\documentclass[prd,twocolumn,showpacs,floats,floatfix,nofootinbib]{revtex4}
\usepackage{dcolumn} 
\usepackage{amssymb}
\usepackage{natbib}
\usepackage{bm} 
\usepackage[pdftex]{graphicx}
\usepackage{amsfonts}
\usepackage{txfonts} 
\usepackage{epstopdf}
\newcommand{\be}{\begin{equation}}
\newcommand{\ee}{\end{equation}}
\newcommand{\bea}{\begin{eqnarray}}
\newcommand{\eea}{\end{eqnarray}}
\newcommand{\beq}{\begin{equation}}
\newcommand{\eeq}{\end{equation}}
\newcommand{\beqar}{\begin{eqnarray}}
\newcommand{\eeqar}{\end{eqnarray}}
\newcommand{\beqars}{\begin{eqnarray*}}
\newcommand{\eeqars}{\end{eqnarray*}}
\newcommand{\bc}{\begin{center}}
\newcommand{\ec}{\end{center}}
\newcommand{\ben}{\begin{enumerate}}
\newcommand{\een}{\end{enumerate}}
\newcommand{\bit}{\begin{itemize}}
\newcommand{\eit}{\end{itemize}}
\newcommand{\bw}{\begin{widetext}}
\newcommand{\ew}{\end{widetext}}
\newcommand{\ex}{\mbox{e}}
\newcommand{\dd}{\mbox{d}}
\def\spose#1{\hbox to 0pt{#1\hss}}

\def\lta{\mathrel{\spose{\lower 3pt\hbox{$\mathchar"218$}}
\raise 2.0pt\hbox{$\mathchar"13C$}}}
\def\gta{\mathrel{\spose{\lower 3pt\hbox{$\mathchar"218$}}
\raise 2.0pt\hbox{$\mathchar"13E$}}}

\def\setR{\mathbb{R}}
\def\setC{\mathbb{C}}

\newcommand{\Hu}{{\cal H}}
\newcommand{\Ka}{{\cal K}}

\newcommand{\GN}{G_{_\mathrm{N}}}
\newcommand{\mP}{m_{_\mathrm{Pl}}}

\begin{document}
\title{Non singular bounce in modified gravity}

\author{L. Raul Abramo}
\email{abramo@fma.if.usp.br}
\affiliation{Instituto de F\'isica, Universidade de S\~ao Paulo, CP 66318, CEP 05314-070,
S\~ao Paulo, Brazil}

\author{Patrick Peter}
\email{peter@iap.fr}
\affiliation{${\cal G}\setR\varepsilon\setC{\cal O}$ --
Institut d'Astrophysique de
Paris, UMR7095 CNRS, Universit\'e Pierre \& Marie Curie, 98 bis
boulevard Arago, 75014 Paris, France}

\author{Ivan Yasuda}
\email{yasuda@fma.if.usp.br}
\affiliation{Instituto de F\'isica, Universidade de S\~ao Paulo, CP 66318, CEP 05314-070,
S\~ao Paulo, Brazil}

\begin{abstract} 
 We investigate bouncing solutions in the framework of the non-singular 
 gravity model of Brandenberger, Mukhanov and Sornborger. We
 show that a spatially flat universe filled with ordinary matter
 undergoing a phase of contraction reaches a stage of minimal
 expansion factor before bouncing in a regular way to reach the expanding phase. 
 The expansion can be connected to the usual radiation- and matter-dominated 
 epochs before reaching a final expanding de Sitter
 phase. In general relativity (GR), a bounce can only take place
 provided that the spatial sections are positively curved, a fact that has
 been shown to translate into a constraint on the characteristic duration of the
 bounce. In our model, on the other hand, a bounce can occur also 
 in the absence of spatial curvature, which means that the timescale for the 
 bounce can be made arbitrarily short or long. The implication is that constraints 
 on the bounce characteristic time obtained in GR rely heavily on the 
 assumed theory of gravity.  Although the model we investigate is fourth order
 in the derivatives of the metric (and therefore
 unstable vis-\`a-vis the perturbations), this generic bounce dynamics 
 should extend to string-motivated non singular models which can 
 accommodate a spatially flat bounce.
\end{abstract} 
\maketitle

\section{Introduction}

Observations, especially those of the CMB by WMAP \cite{Komatsu:2008hk},
strongly suggest the occurrence of a primordial inflationary period
\cite{Guth,Linde-PL,Albrecht-Steinhardt}.  Inflation not only provides
an explanation for the homogeneity, flatness and horizon problems of
the standard hot big bang cosmology, but it also offers a consistent
mechanism under which metric fluctuations are stretched beyond the
Hubble radius with a nearly invariant power spectrum. However, in
spite of its successes, the theory of inflation does not solve the
problem of the initial singularity. Although the weak energy condition
is likely to be violated in such models, Borde, Guth and Vilenkin have
shown \cite{Borde-Vilenkin-PRD} that inflating spacetimes are in
general geodesically incomplete to the past
\cite{Borde-Vilenkin-PRL,Borde-Guth-Vilenkin-PRL}.

Mostly inspired by the string motivated pre-big bang scenarios
\cite{Gasperini-Veneziano-PR,Lidsey-PR}, bouncing models
\cite{Murphy,Melnikov,Fabris-Peter,Martin-Peter,Martin-Peter-Pinto-Neto-PRD},
i.e., models in which the universe undergoes a phase of contraction
followed by expansion, have been proposed as alternatives to the
inflationary paradigm. A bounce could solve the flatness problem of
standard cosmology if the contracting phase lasted much longer than
the expanding one, and it could also solve the homogeneity problem by
making the past light cone large so that thermalization could occur.

The problem of the initial singularity, however, is still generic. In
the pre-big bang scenario the cosmological field equations exhibit a
new symmetry, the scale factor duality, which maps the pre-big bang,
contracting dilaton dominated era (in the Einstein frame) to the usual
Friedmann-Robertson-Walker cosmology (post-big bang phase)
\cite{Veneziano-arxiv}. Nonetheless, it has been shown that the two
branches cannot be connected to each other smoothly
\cite{Brustein-Veneziano}. This means that, while the pre-big bang era
has a future singularity, the post-big bang phase emerges from a past
singularity.

The issue of how metric perturbations could be affected by the bounce
has also been addressed, mostly in the framework of General Relativity
(GR). The low energy approach represented by GR is natural if we
consider the fact that high energy corrections coming from, say,
string theory, are negligible before and after the bounce. Of course,
this is no longer the case if we want to fully describe the bounce
mechanism itself. This uncomfortable situation has led to the
postulate that, in analogy to other short transitions in standard
cosmology, such as pre-heating \cite{finelli-brandenberger-PRL} or
radiation to matter dominated epochs
\cite{Mukhanov-Brandenberger-Perturbations}, the time scale of the
bounce is such as to permit fluctuations to evolve through it in a
scale-invariant way \cite{Gasperini-Veneziano-PR,Durrer}.  In the GR
framework, however, this assumption is far from being generic.  On the
contrary, it has been shown \cite{Martin-Peter-PRL} that large
wavelengths do suffer the influence of a such cosmological transition.
Hence the need to understand the way the characteristic bounce time is
constrained (or not) by the field equations.

In addition, because GR forbids the bounce to occur as long as the
null energy condition holds, setting up such an evolution for the
scale factor can be a challenge. For instance, in a scenario with a
single scalar field and positive spatial curvature
\cite{Martin-Peter}, the spectrum in the large scale limit exhibits
$k$-mode mixing. Models without spatial curvature but with a generic
scalar field -- or $k$-essence field \cite{Armendariz-PRD} -- as
matter content have also being built \cite{Abramo-Peter}. In that case,
besides the fact that physical observables are still affected by the
bounce, those scenarios all lie in the {\sl phantom} sector. Thus, in
order to connect the $k$-bounce to an expanding radiation era, a decay
mechanism similar to pre-heating would be necessary.

In this work, we neither look for regularizations of the pre-big bang
scenario nor for some other matter/curvature configuration in the
classical realm. Rather, we will adopt an alternative approach,
focused on a modified, higher order derivative gravity model proposed
in \cite{Mukhanov-Brandenberger-PRL,Mukhanov-Brandenberger-PRD}.  In
this approach, an effective action for gravity is constructed in such
a way that all curvature invariants are limited. This is done via
non-dynamical Lagrange's multipliers whose potentials ensure that the
theory approaches Einstein's gravity at low curvature and that all
solutions are well behaved at high curvature.

In a subsequent paper \cite{Brandenberger-Easther}, as an attempt to
regulate singularities in the pre-big bang cosmology, Brandenberger,
Easther and Maia studied a non singular dilaton cosmology in the
framework of the model presented in
\cite{Mukhanov-Brandenberger-PRL,Mukhanov-Brandenberger-PRD}.  They
found solutions corresponding to a contracting, dilaton-dominated
universe which evolves toward a bounce and emerges as a Friedmann
universe. Here we propose another solution: we will consider a
homogeneous and isotropic universe filled with ordinary radiation.
The choice of curvature invariants will dictate the dynamics for the
scale factor, which develops from a regular bounce to a Friedmann
expanding universe, ending up with a quasi de Sitter expansion phase
which could mimic the present acceleration of the universe. We also
address the problem of the duration of the bounce in this model, and
show that, as opposed to the classical treatment \cite{Martin-Peter},
it turns out to be completely unconstrained by the field equations.

Extensions of GR involving higher powers of curvature invariants are
well justified, as calculations of one-loop divergences in quantum
gravity generate terms proportional to $R^{2}$, $R_{\mu\nu}^{2}$ and
$R_{\mu\nu\rho\sigma}^{2}$ \cite{Hooft,Deser-Nieuwenhuizen-PRD}.  As
shown by Stelle in $1977$ \cite{Stelle}, although such actions can
lead to a renormalizable theory, they all have a shortcoming, namely,
the presence of ghosts -- degrees of freedom with negative kinetic
energy \cite{Chiba-JCAP,Nunez}. This fact makes the theory highly
unstable, in the sense that the vacuum (empty) state can decay into a
collection of both positive and negative energy states.  Also, at the
classical level, one should expect that such instability will lead to
growing gravitational perturbations carrying both positive and
negative energy modes. This phenomenon is known in the literature as
Ostrogradsky's instability \cite{Woodard,Bruneton-Esposito}.  It is in
close relationship with the fact that the Hamiltonian, due to the
presence in the Lagrangian of derivative terms of order greater than
one, is unbounded from below.

In this work we will consider a simple quadratic theory, namely, a
theory for which the higher order derivative terms appear only
linearly. As a consequence, the background field equations will be of
second order at the background level.  This fact, however, does not
suffice to prove stability, as the perturbations would still possess
higher-order equations of motion.

This paper is organized as follows. In the next section, we outline
the model and write the full equations of motion, in terms of both the
cosmic and conformal times. In section III we show how the universe
can pass through a non-singular bounce and then connect to the usual
expanding phases of standard cosmology. The final phase approaches
asymptotically to an ever-expanding de Sitter period.  At the end of
section III we show how the time duration of the bounce can be made
arbitrarily short. We conclude in section IV.

\section{The Model}

Our goal is to investigate bouncing cosmological solutions in a
non-singular, higher order derivative gravity proposed in
Refs.~\citep{Mukhanov-Brandenberger-PRL,Mukhanov-Brandenberger-PRD}.
In this model combinations of the Riemann tensor are introduced into
the action via non-dynamical Lagrange multipliers. The potentials
associated to each of the multipliers ensure that the theory
approaches the Einstein limit at low spacetime curvature, and that the
solution to the field equations is non-singular (typically, a de
Sitter solution) at high curvature. More generally, for the
gravitational action, we have
\begin{equation}
 S_\mathrm{grav}  = \frac{1}{16\pi\GN} \int
 \left[ R+\sum_{i=1}^N \varphi_i I^{(i)}-V
   \left(\varphi_{i}\right)\right]\sqrt{-g}\,\dd^{4}x,
\label{action}
\end{equation}
where $\GN^{-1}=8\pi\mP^2$ is the Newton constant defining the Planck
mass $\mP$, $\varphi_{i}$ ($i=1,\cdots,N$) are Lagrange multipliers
depending on space-time coordinates, and $I^{(i)}$ are functions of
the curvature invariants
\begin{equation}
 I^{(i)}=I^{(i)} \left( R,R_{\mu\nu}R^{\mu\nu},
   R_{\mu\nu\alpha\beta}R^{\mu\nu\alpha\beta},...\right),
\label{Ii}
\end{equation}
which we would like to limit.

In order to understand how these limits are implemented, let us
restrict ourselves to the case where a single field
$\varphi_{1}\equiv\varphi$ is present, and only one invariant,
$I^{(1)}\equiv I$, is limited.  Variation of the action (\ref{action})
with respect to $\varphi$ then provides the constraint equation
\begin{equation}
I - \frac{\dd V}{\dd\varphi} = 0.
\label{constraint}
\end{equation}
At low curvature, we demand that $\varphi$ be small and the theory to
approach GR. Hence the action in this limit should be approximated by
\begin{equation}
 \lim_{\varphi\to 0} S_\mathrm{grav}=  \frac{1}{16\pi\GN} \int \left[ R+
   \mathcal{O}\left(\varphi,\varphi^{2},\cdots \right)\right]\sqrt{-g}\,\dd^{4}x,
\label{GRlim}
\end{equation}
which implies, from Eq.~(\ref{constraint}), that the potential should
behave as
\begin{equation}
 \lim_{\varphi\to 0} V\left(\varphi\right)\sim\varphi^{2}+\varphi^{3}+\cdots.
\label{condition_a}
\end{equation}

On the other hand, at high curvature, which we take to mean
$|\varphi|\gg 1$, the correction term for the Einstein-Hilbert action
becomes important. The potential must then be chosen in such a way
that the solution of the field equations approaches the de Sitter
solution, thereby effectively limiting the curvature.  From
Eq.~(\ref{constraint}), we see that this requirement can be fulfilled
provided that
\begin{eqnarray}
 \lim_{\varphi\to \infty} V\left(\varphi\right) & = & 2\Lambda,\nonumber \\
 \lim_{\varphi\to \infty} I\left(R,R_{\mu\nu}^{2},...\right) & =& 0,
\label{condition_b}
\end{eqnarray}
where $\Lambda$ is a constant.  Thus, the action in the high curvature
regime can be written as
\begin{equation}
 \lim_{\varphi\to \infty} S_\mathrm{grav} = 
 \frac{1}{16\pi\GN} \int\left( R-2\Lambda\right) 
 \sqrt{-g}\,\dd^{4}x.
\label{limitInfy}
\end{equation}
We thus recover the Einstein-Hilbert action with a cosmological
constant term coming from the potential.  A detailed discussion of
this de Sitter limit can be found in
\cite{Mukhanov-Brandenberger-PRD}.

Here we are mainly interested in the particular case where the
universe, otherwise filled with a perfect fluid such as, e.g., dust or
radiation, is originally undergoing a phase of contraction where the
spacetime curvature increases. Since the curvature is now limited, the
Universe must reach a minimum radius, at which point it should bounce
and subsequently expand.

Such a cosmology should be realizable with a convenient choice of
curvature invariants. Assuming a Friedmann-Lema\^{\i}tre-Robertson
Walker (FLRW) metric of the form
\begin{equation}
 \dd s^{2}=-\dd t^{2}+a^{2}\left(t\right)\left[ 
   \frac{\dd r^2}{1-\Ka r^2} +r^2\left(
     \dd\theta^2 +\sin^2\theta \dd\phi^2\right) \right],
\label{metric}
\end{equation}
we find that singularity-free second order equations can be obtained
with the curvature invariant
\begin{equation}
 I\left(R,R_{\mu\nu}R^{\mu\nu}\right)\equiv 
 R-\sqrt{3\left(4R_{\mu\nu}R^{\mu\nu}-R^{2}\right)} =12\left(\dot{H}+H^{2}\right),
 \label{I}
\end{equation}
where the Hubble expansion rate is $H\equiv\dot a/a$, provided that
the potential obeys the twin conditions that $V(0)=0$, $V'(0)=0$ and
$V'(\pm \infty) \rightarrow$ constant.

\begin{figure}
 \hskip-5mm\includegraphics[scale=.5]{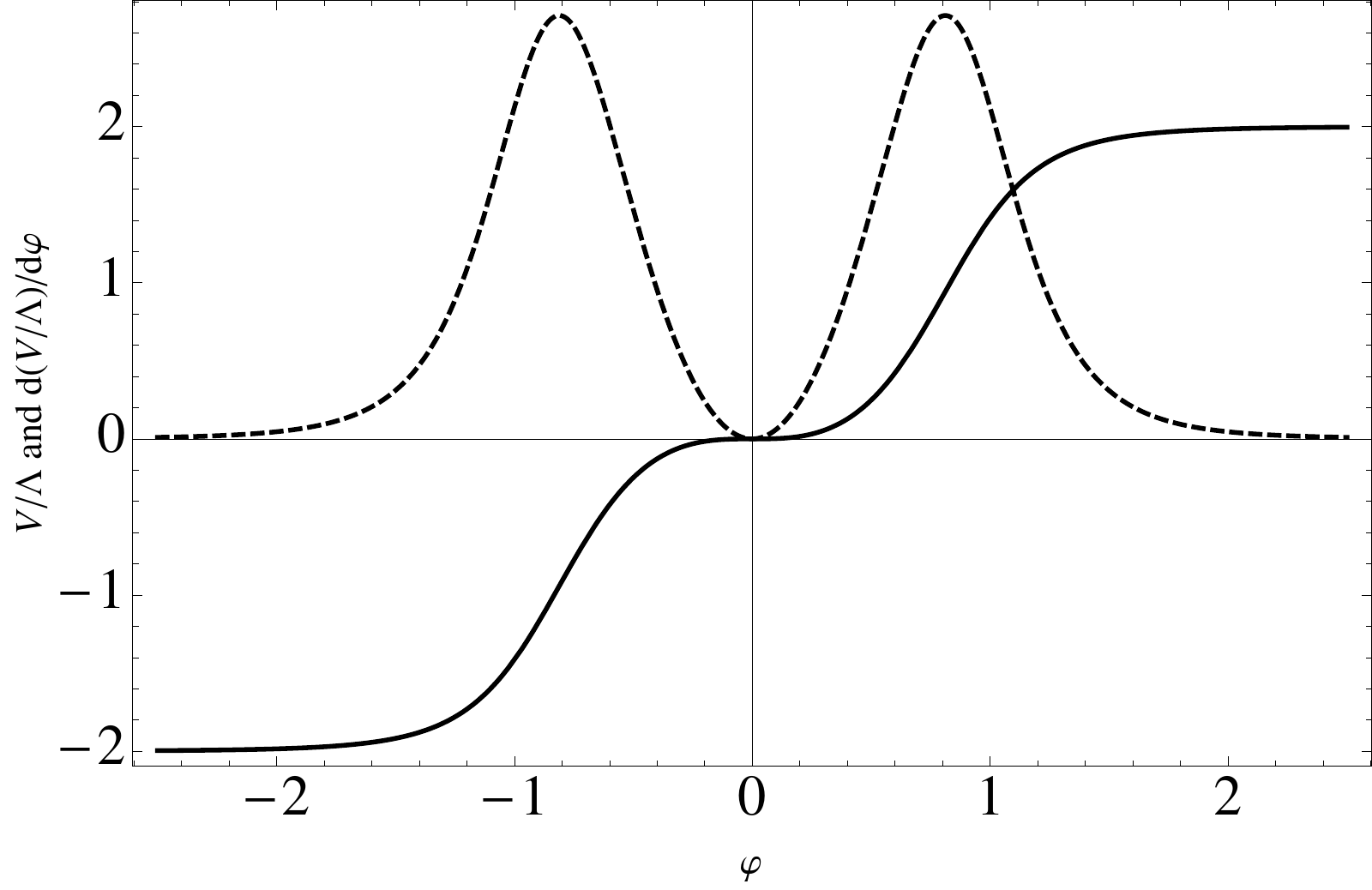}
 \caption{ Scalar field potential $V(\varphi)$ (full line)
   and its derivative (dashed line), functions of the scalar field
   $\varphi$, as used for the modified gravity field equations.
   \label{fig:V_phi}}
\end{figure}

One simple example of a potential that satisfies the conditions spelled above 
would be
\begin{equation}
 V\left(\varphi\right) = \frac{2\Lambda\varphi^{3}}{\sqrt{1+\varphi^{6}}},
\label{invariant}
\end{equation}
and it is represented on Fig.~\ref{fig:V_phi}. Inserting this
potential into Eq.~(\ref{constraint}), we obtain
\begin{equation}
 \dot{H}+H^{2}=\frac{1}{12}\frac{\dd V}{\dd\varphi} = 
 \frac{\Lambda\varphi^{2}}{2\left(1+\varphi^{6}\right)^{3/2}}.
\label{constraint_b}
\end{equation}
Eq.~(\ref{constraint_b}) tells us that, at the bounce where $H=0$, we
can have $\dot{H}>0$ provided we choose a positive cosmological
constant $\Lambda$. Besides, we want $H$ to reach a maximum value
after the bounce, after which it should decrease again -- we put this
extra ingredient in order to illustrate the connection between the
bounce era and the usual cosmological setting of a decelerating
($\dot{H}<0$) Universe. In the next section we will show that this
feature is realized for a range of initial conditions.

The complete gravitational action we choose is therefore
\begin{equation}
 S_\mathrm{grav} = \frac{1}{16\pi\GN} \int \sqrt{-g}\, 
 \mathcal{L}_\mathrm{grav}\,\dd^4x,
\label{final_action}
\end{equation}
where
\begin{equation}
 \mathcal{L}_\mathrm{grav}= R + \varphi \left[
   R-\sqrt{3\left(4R_{\mu\nu}R^{\mu\nu}-R^{2}\right)}\right] -V\left(\varphi\right).
\label{final_lag}
\end{equation}
Variation of the action (\ref{final_action}) with respect to the
metric then leads to the modified Einstein equations
\begin{eqnarray}
 R_{\mu\nu} & - & \frac{1}{2}g_{\mu\nu}R-g_{\mu\nu}\frac{\varphi I}{2}+
 \varphi I_{R}R_{\mu\nu}-\left(\varphi I_{R}\right)_{;\nu;\mu}
 +g_{\mu\nu}\left(\varphi I_{R}\right)_{;\rho}^{;\rho}\nonumber \\
 & + & 2\varphi I_{P}R_{\mu\rho}R_{\nu}^{\rho}+
 \left(\varphi I_{P}R_{\mu\nu}\right)_{;\rho}^{;\rho}
 +g_{\mu\nu}\left(\varphi I_{P}R^{\rho\lambda}\right)_{;\lambda\rho}
 \nonumber \\
 & - & 2\left[\varphi I_{P}R_{(\mu}^{\lambda}\delta_{\nu)}^{\rho}\right]_{;\rho;\lambda}
 +\frac12 g_{\mu\nu}V\left(\varphi\right)= 8\pi\GN T_{\mu\nu},
\label{EoM}
\end{eqnarray}
where we have defined $I_{R}\equiv\partial I/\partial R$ and
$I_{P}\equiv\partial I/\partial\left(R_{\mu\nu}R^{\mu\nu}\right)$; in
Eq.~(\ref{EoM}), we also added a matter component whose stress-energy
tensor is $T_{\mu\nu}$.  These equations should be solved together
with Eq.~(\ref{constraint_b}).

Eqs.~(\ref{EoM}) are, in general, of fourth order in the metric
variables, and therefore subject to Ostrogradsky's
instability. However, in the special case of the maximally symmetric
FLRW metric (\ref{metric}), and for the particular choice of the
invariant $I$ of Eq.~(\ref{I}), we find that the second derivative of
the scale factor appears only linearly in the Lagrangian
(\ref{final_lag}).  As a result, the background equations of motion stemming from
(\ref{EoM}) are second order. Indeed, for the time-time component we
obtain the modified Friedmann equation
\begin{equation}
 H^{2}\left(1+4\varphi\right) + 2H \dot{\varphi} +\frac{\Ka}{a^2}
 = \frac{V}{6}+\frac{8\pi\GN}{3}\rho,
\label{time-time}
\end{equation}
with $\rho$ the matter energy density of the perfect fluid.  For the
spatial components the equations again lead to a generalization of the
usual GR case, namely
\begin{equation}
 \ddot{\varphi} +4 H\dot{\varphi} + \left(\dot{H}+
   \frac32 H^{2}\right) \left(1+4\varphi\right) +\frac{\Ka}{2a^2}
 = \frac{V}{12} - 4\pi\GN p,
\label{space-space}
\end{equation}
where $p$ is the matter pressure. This last equation can also be
derived from Eq.~(\ref{time-time}) and the matter stress-energy tensor
conservation.

A simple way of obtaining these equations is by setting
$g_{00}=-N^2(\tau)$, $\tau$ being a general time coordinate. This
transforms the gravitational action into
\begin{equation}
 S_\mathrm{grav} = \frac{3}{8\pi\GN}\int \sqrt{\gamma} \,\dd^4 x\, 
 L\left(N,\varphi\right),
\label{ADMact}
\end{equation}
where $\sqrt{\gamma} = r^2\sin\theta/\sqrt{1-\Ka r^2}$ is the square
root of the determinant of the metric on the spatial sections, and
\begin{equation}
 L = a^3\left\{ N \left( \frac{\Ka}{a^2}
     -\frac{V}{6} \right) -\frac{1}{N} 
   \left[ \left(\frac{\dd a}{a\dd\tau}\right)^2 +2 \frac{\dd a}{a\dd\tau}
     \left( \frac{\dd\varphi}{\dd\tau}+2 
       \frac{\dd a}{a\dd\tau} \varphi\right) \right] \right\}.
\label{ADMLag}
\end{equation}
Varying Eq.~(\ref{ADMact}) with respect to $\varphi$ and $N$ yields
Eqs.~(\ref{constraint_b}) and (\ref{time-time}) respectively once one
fixes the lapse function $N$ to unity, i.e., once one identifies
$\tau$ with the cosmic time $t$.

For later use it is convenient to rewrite the equations of motion in
terms of the conformal time $\eta$, defined in the usual way as
$\dd\eta=\dd t/a$. Denoting by a prime the derivative with respect to
conformal time, $f'\equiv \dd f/\dd\eta$, we rewrite the constraint
equation as
\begin{equation}
 \Hu' = \frac{ a^2}{12} \frac{\dd V}{\dd\varphi},
\label{VincConf}
\end{equation}
and the generalized Friedmann equation as
\begin{equation}
 \Hu^2\left( 1 + 4 \varphi \right) + 2 \Hu \varphi' + \Ka 
 = a^2 \left(\frac{V}{6}+\frac{8\pi\GN}{3}\rho \right).
\label{FriedConf}
\end{equation}
These last two equations can also be derived from variations of
(\ref{ADMact}), setting $\tau=\eta$, i.e. $N=a$. Together with matter
energy-momentum conservation
\begin{equation}
 \rho'+3\Hu \left(\rho+p\right)=0,
\label{ConsConf}
\end{equation}
these equations can be combined to yield
\begin{equation}
 \Hu'\left(1+4\varphi\right)+\varphi''+2\Hu \varphi'=a^2 \left[ \frac{V}{6}
   -\frac{4\pi\GN}{3}\left( \rho+3p\right)\right].
\label{dotHgen}
\end{equation}
Eqs.~(\ref{VincConf}), (\ref{FriedConf}) and (\ref{ConsConf}) provide
a self-contained description for the bounce, to which we shall now turn to.

\section{Bouncing universe}

Before studying the different cases of interest, let us use the
cosmological constant $\Lambda$ to define a more convenient time
variable $\tau = \sqrt{\Lambda} t$.  Rescaling the Hubble parameter,
the scale factor and the energy density through
\begin{equation}
 H=\sqrt{\Lambda} h(\tau), \ \ \ \ \tilde a = \Lambda a
 \ \ \ \hbox{and} \ \ \ \varepsilon(\tau) = \frac{8\pi\GN}{3}\frac{\rho}{\Lambda},
\label{rescal}
\end{equation}
we obtain a set of dimensionless equations as
\begin{equation}
 \frac{\dd\varphi}{\dd\tau} = \frac{\mathcal{V}}{12 h} -\frac{1}{2} h\left(1+4\varphi\right)
 - \frac{\Ka}{2 \tilde a^2h}+\frac{\varepsilon}{2h},
\label{phi_0dim}
\end{equation}
and
\begin{equation}
 \frac{\dd h}{\dd\tau} = \frac{1}{12}\frac{\dd\mathcal{V}}{\dd\varphi}-h^2,
\label{h_0dim}
\end{equation}
where the potential is now $\mathcal{V} = V/\Lambda$.

The matter stress-energy conservation then reads
\begin{equation}
\frac{\dd\varepsilon}{\dd\tau} + 3 h \left( \varepsilon + \mathcal{P} \right) =0,
\label{rho_0dim}
\end{equation}
where the pressure $p$ is modified in the same way as the energy
density through $\mathcal{P} = 8\pi\GN p/(3\Lambda)$.  These
rescalings also apply for the conformal time equations, with $\Hu$
unchanged.

\subsection{Dynamics of the field equations}

We begin the section with a dynamical analysis of the vacuum equations
with zero spatial curvature ($\Ka=0$), in the case of the potential of
Eq.~(\ref{invariant}).  We find that the system
\begin{eqnarray}
 \frac{\dd h}{\dd\tau} & = & \frac{\varphi^{2}}{2\left(1+\varphi^{6}\right)^{3/2}} -h^{2},
 \label{h_eq}\\
 \frac{\dd\varphi}{\dd\tau} & = & \frac{\varphi^{3}}{6h\sqrt{1+\varphi^{6}}}
 -\frac{1}{2} h \left(1+4\varphi\right)
\label{varphi_eq},
\end{eqnarray}
has six critical points (the number of crossings of the curves defined
by $\dd h/\dd \tau=0$ and $\dd\varphi/\dd \tau=0$) -- see Fig. 1.  One
of these critical points, at $(\varphi,h) \simeq (-0.3, 0.21)$, is an
asymptotically stable point -- an attractor. This means that
trajectories come to an end (at $t \rightarrow \infty$) at this point,
which is hence a de Sitter attractor.

\begin{figure}
 \hskip-5mm\includegraphics[scale=0.55]{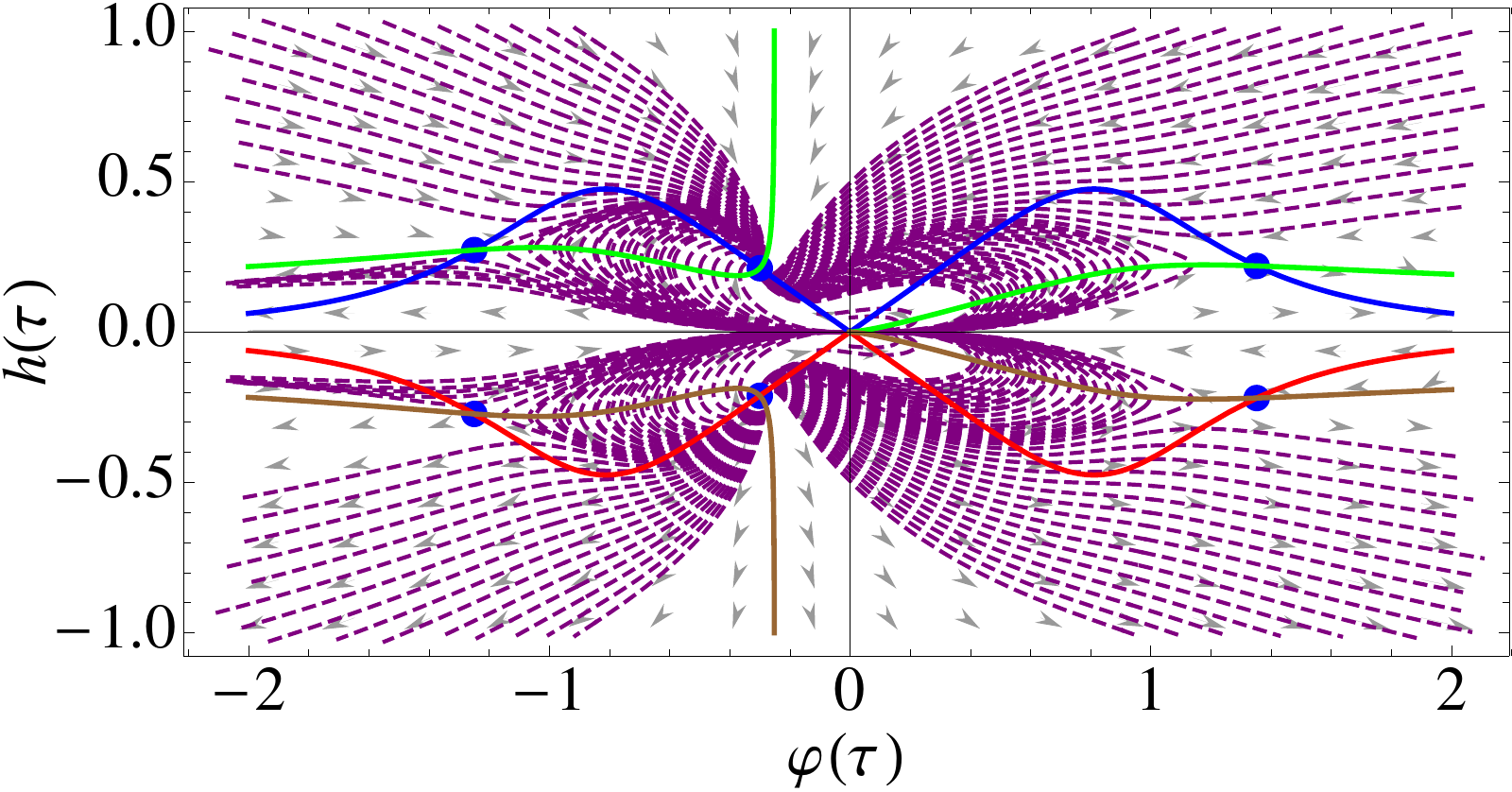}
\caption{Phase-space diagram in vacuum. The critical points are shown
as filled circles, and are found at the intersections of 
the curves corresponding to $\dd H/\dd \tau=0$ and
$\dd\varphi/\dd \tau=0$ (solid lines). The dashed lines show some
particular solutions on this phase space.
The arrows depict the arrow of time.
\label{fig:PS-diagram-vacuum}}
\end{figure}

The point $(\varphi,h) \simeq(-0.3, -0.21 )$, on the other hand, is an
unstable spiral point -- where trajectories emerge from at
$t\rightarrow -\infty$. It corresponds to an anti-attractor, in this
case, a contracting de Sitter solution.  The other critical points, to
the left and right of Fig. 1, namely, $(\varphi,h) \simeq (-1.25,
-0.27)$, $(-1.25, 0.27)$, $(1.35, -0.22)$ and $(1.35, 0.22)$, are all
saddle points.  Some particular solutions for this system are sketched
in Fig. \ref{fig:PS-diagram-vacuum}.

The most important feature of this phase diagram, for us, is that the
origin $(\varphi=0,H=0)$ is not a critical point -- it corresponds to
the crossing of the two branches of the curves $\dd H/\dd t=0$.  In
fact, trajectories coming from below ($H<0$) will necessarily have to
go through the origin in a finite interval of time.  But the
trajectories from below all cross the same point (the origin), which
means that (at least for the autonomous system we considered thus far)
this point is somehow singular.  This means, in our case, that the
trajectories passing through the origin will experiment a
discontinuous change in the rate of change of $\varphi$ -- in other
words, one cannot predict the value of $\dd\varphi/\dd t$ for
trajectories that emerge from the origin at $H = 0$.

To see that this is the case, we look for solutions near the origin
$\left(\varphi=0,H=0\right)$ of the phase plane. In this limit, the
dynamical system can be approximated by
\begin{equation}
 \frac{\dd h}{\dd \varphi}=h-\frac{\varphi^{2}}{h}+
 \mathcal{O}\left(h^{2},\varphi^{2},h\varphi\right).
\label{approx_ps-eq-vacuum}
\end{equation}
We can rewrite the above equation in terms of $h^{2}$, and then find a
first integral of the resulting differential equation. The final
result is that
\begin{equation}
 h\left(\varphi\right) \simeq \pm\frac{1}{4}
 \sqrt{A \ex^{4\varphi}+1+4\varphi\left(1+2\varphi\right)},
\label{approx_ps_H_sol-vacuum}
\end{equation}
where $A$ is an arbitrary integration constant. This means that the 2
dimensional dynamical system is still Lipschitz continuous at the
origin (see, e.g., Ref. \cite{Jeffreys}, pg. 53), which indicates that
the 2 dimensional trajectories can be seen as projections of
trajectories in a higher-dimension dynamical phase space. Hence,
Eq.~(\ref{approx_ps_H_sol-vacuum}) indicates that although the two
branches ($H>0$ and $H<0$) cannot be simply connected to each other in
the case of the 2 dimensional phase space, any enlargement of that
phase space will lift the problematic degeneracies.

The saddle points divide the phase diagram in different regions
through their separatrices (not shown in
Fig. \ref{fig:PS-diagram-vacuum}, although they can be inferred from
the particular solutions.)  In particular, the separatrices in the
third and fourth quadrants explain why there are solutions which never
reach the $H=0$ axis. This is a general feature, in the sense that it
does not depend on the matter content: indeed, it
follows directly from Eq.~(\ref{constraint_b}), which is
independent of the matter content for the constrained theory we are
considering. However, the evolution of the right-hand side of
Eq.~(\ref{constraint_b}) being coupled to the matter content through
Eq.~(\ref{time-time}), one expects a similar general behavior in the
Hubble parameter for universes with any type of matter content, but
with only minor differences in the details of the phase diagram.

Hence, we conclude that a bounce in this constrained gravity theory
does seem quite generic, but unfortunately the contracting and
expanding branches cannot be connected in a continuous (unambiguous)
way to each other, because the dynamical system is two-dimensional.
However, as we will show next, the introduction of even a minuscule
amount spatial curvature $\Ka$, or of any kind of matter, both opening
up the phase space into a third dimension, is enough to regularize
this bounce, so that every physical variable in the system remains
finite and no singularity occurs.

\subsection{Spatial curvature}

By inspection of the phase diagram of
Fig. \ref{fig:PS-diagram-vacuum}, we can see that for every trajectory
in the lower plane ($h<0$) there is a similar trajectory in the upper
plane. The only problem is that apparently all these particular
solutions cross at the origin $(\varphi=0,h=0)$, which, sadly for the
2 dimensional dynamical system we considered in the previous Section,
cannot make sense. As it turns out, spatial curvature is one way to
alleviate this problem.

\begin{figure*}[ht]
\hskip-5mm
\includegraphics[scale=0.35]{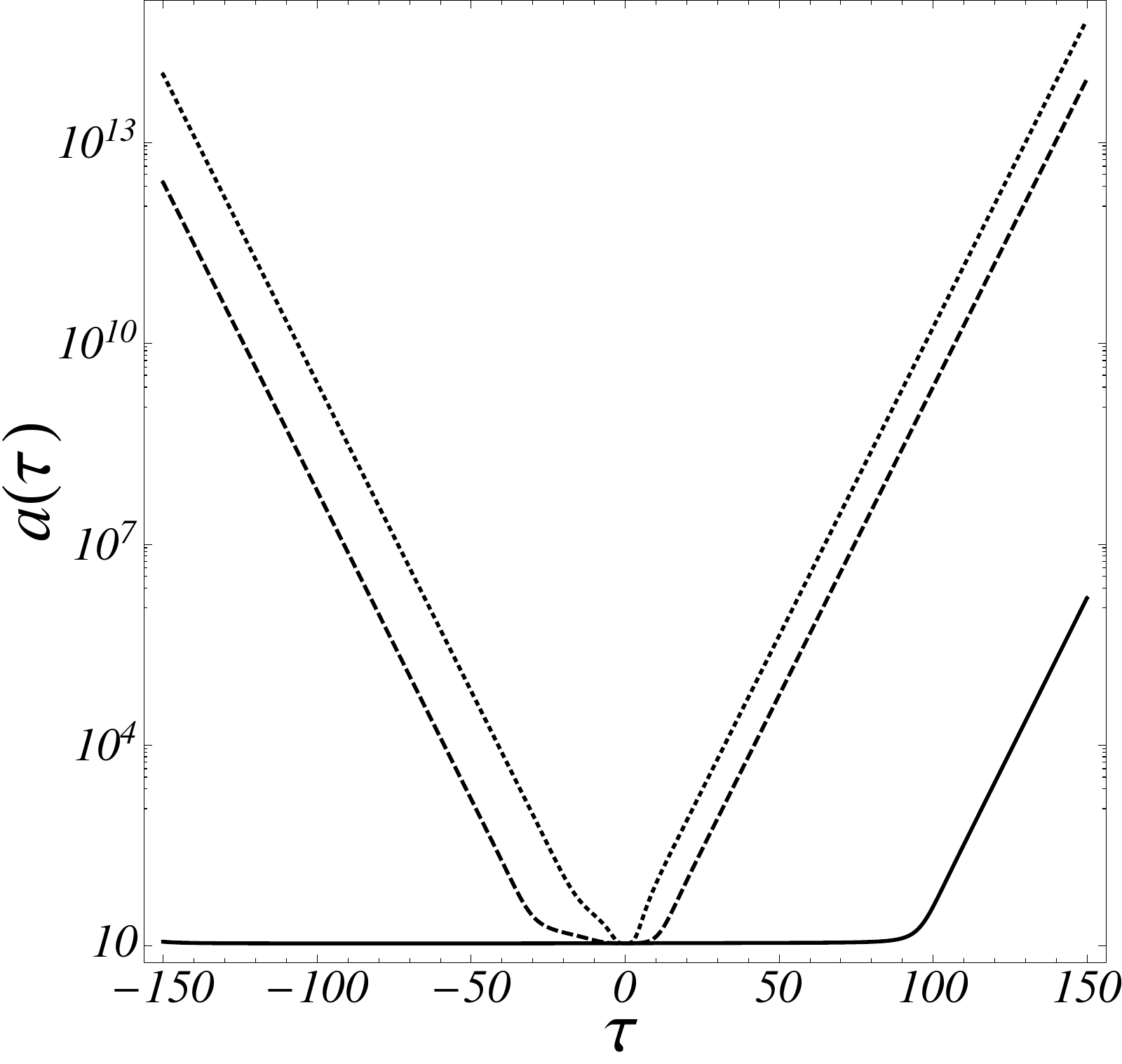}
\hskip2mm
\includegraphics[scale=0.35]{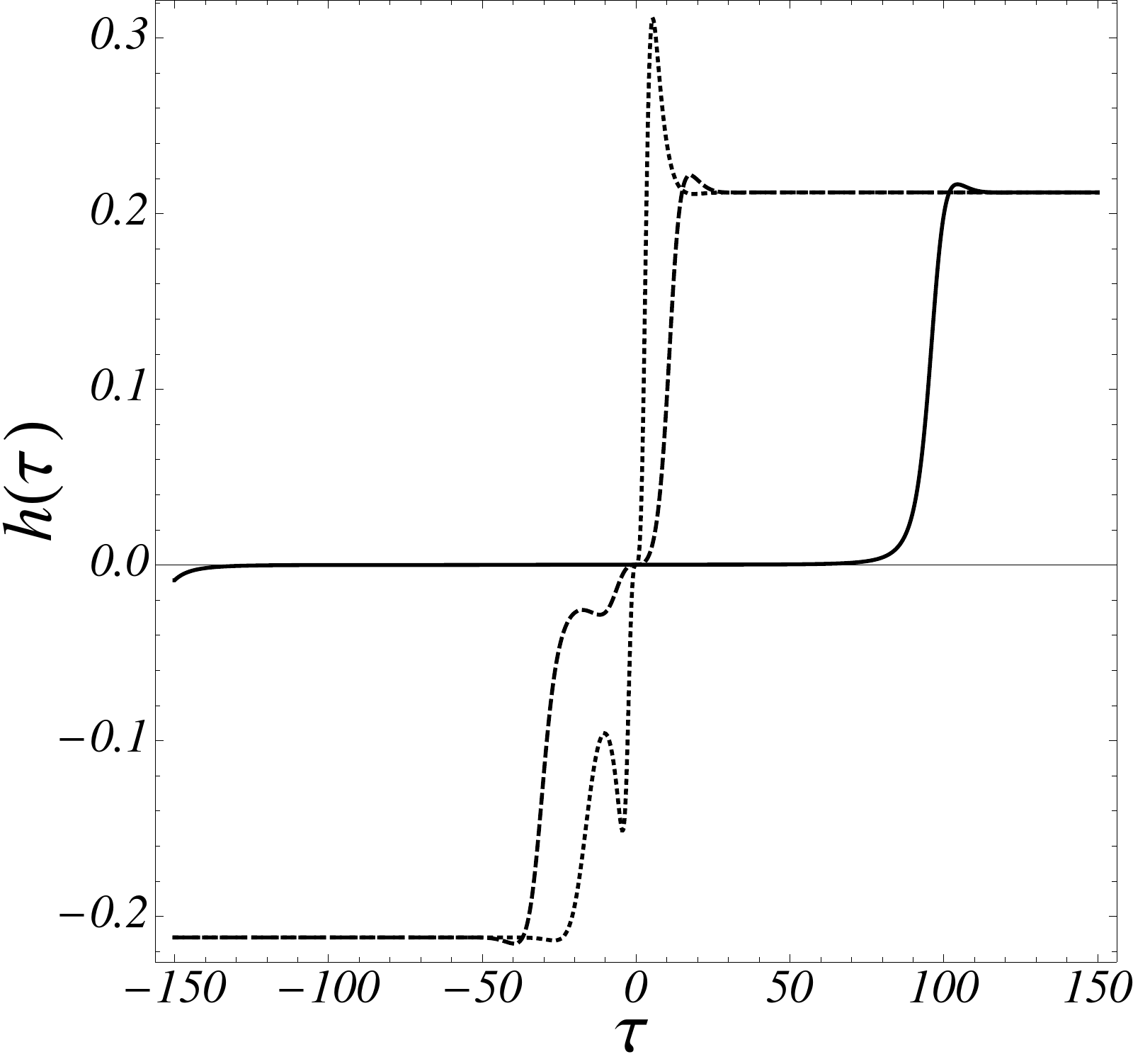}
\hskip2mm
\includegraphics[scale=0.35]{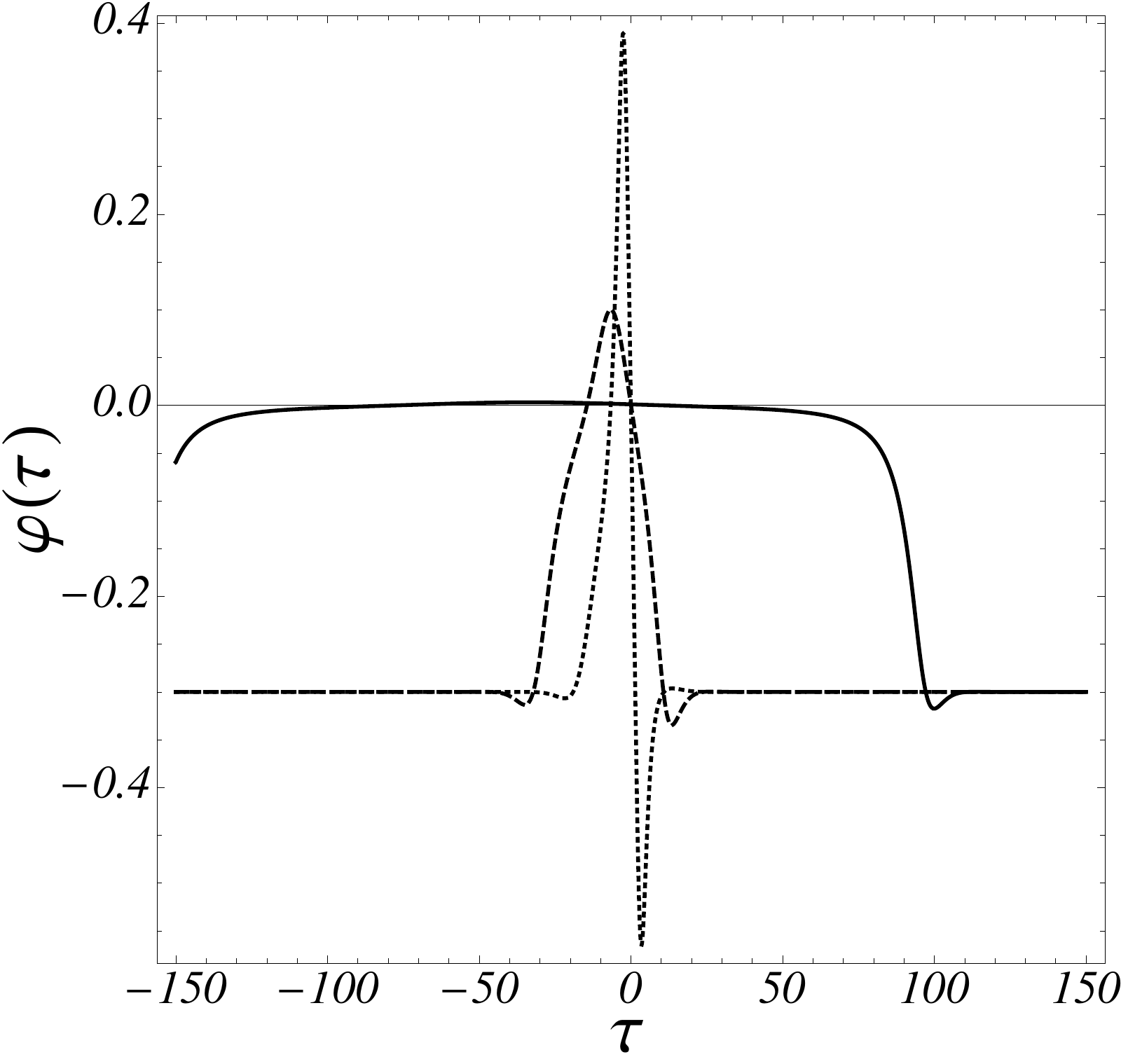}
\caption{Scale factor $a(\tau)$ (left panel), reduced Hubble
parameter $h(\tau)$ (center) and scalar field
$\varphi(\tau)$ (right panel) as a function of time for 
almost vanishing curvature ($\Ka=10^{-9}$, full line) and small but non
vanishing curvature  ($\Ka=10^{-3}$ and $\Ka=10^{-2}$, respectively
dashed and dotted lines). It is interesting to note that the vanishing
curvature limit allows for a so-called emerging Universe model in
which the scale factor is essentially constant for a long period of
time after which expansion (in fact inflation) takes place
spontaneously. Pushing the calculation backwards in time shows that
the vanishing of the Hubble rate is only stable in the actual limit
$\Ka\to 0$: expansion begins after a finite amount of time, but the
amount of past-time during which $h=0$ can only be made infinite
provided $\Ka=0$ strictly; otherwise, the Minkowski state always
originates from a contracting de Sitter phase.
\label{fig:a_tau_K}}
\end{figure*}

Indeed, with even a small amount of spatial curvature, trajectories
coming up from the lower-right corner ($\varphi >0$,
$\dd\varphi/\dd\tau<0$ $h<0$, $\dd h/\dd \tau \gtrsim 0$) emerge from
the origin on the upper left corner ($\varphi <0$, $\dd\varphi/\dd
\tau<0$ $h>0$, $\dd h/\dd \tau \gtrsim 0$), and solutions coming up
from the lower-left corner emerge on the upper-right corner, so that
the particular solutions to the equations of motion are indeed
perfectly continuous. This can only happen because the presence of
spatial curvature turns our original (flat) 2 dimensional phase space
into a 3 dimensional phase space, where $h$ and $a$ are effectively
independent dynamical variables. This means that the trajectories of
Fig. \ref{fig:PS-diagram-vacuum} which cross $\varphi=0$ and $h=0$ can
now do so freely, as the third dimension (the scale factor $a$, shown
on Fig. \ref{fig:a_tau_K}) opens the possibility that the paths do not
cross -- in other words, in the 3 dimensional phase space these trajectories
actually do not intersect.

To see this, consider the dynamical system composed of
Eqs.~(\ref{VincConf}) to (\ref{ConsConf}). The problematic equation,
which leads to the difficulties exposed in the previous section, is
Eq.~(\ref{FriedConf}).  In fact, by taking the limit $\Hu \rightarrow
0$ at the putative bounce we can see that this equation reduces to
\begin{equation}
 \nonumber
 \varphi' \simeq \frac{1}{2\Hu} \left( \frac{a^2 V}{6} - \Ka \right).
\end{equation}
The rate of change of the scale factor is naturally zero at the bounce
($\Hu \rightarrow 0$), hence, if the derivative of the field above is
finite, then $\varphi \rightarrow \varphi_\mathrm{B}$ and $a
\rightarrow a_\mathrm{B}$ at the bounce. This means that
$V(\varphi_\mathrm{B}) = 6 \Ka/a_\mathrm{B}^2$, which then implies,
for the potential given in Eq.~(\ref{invariant}), that
\begin{equation}
 \varphi_\mathrm{B}^6 = \left( \frac{\Lambda^2 
     a_\mathrm{B}^4}{9 \Ka^2} -1 \right)^{-1}.
\label{varphi_B}
\end{equation}
Clearly, as $\Ka \rightarrow 0^\pm$, $\varphi_\mathrm{B} \rightarrow
0$, which means that the physical system itself is perfectly
well-behaved for any nonzero value of $\Ka$.

Hence, we conclude that it is possible to connect the two branches of
the theory (the contracting and expanding phases), provided that there
is any amount of curvature, however small.  Since the limit $\Ka
\rightarrow 0$ is well-behaved, we can do without it entirely, and
join the pieces of the trajectories between the two branches,
constructing zero-curvature bounce models.

It is also interesting to notice that the condition (\ref{varphi_B})
implies that there is a maximum value for the spatial curvature
\begin{equation}
 \left| \Ka_\mathrm{max} \right|  = 
 \frac{\Lambda a_\mathrm{B}^2}{3} = 
 H_\Lambda^2 a_\mathrm{B}^2,
 \label{Kmax}
\end{equation}
which is in fact perfectly legitimate in our constrained theory, since
we limit the spacetime curvature at all times -- and that includes the
time of the bounce itself. These considerations are illustrated on
Fig. \ref{fig:a_tau_K}, showing the time evolution of the scale
factor, the reduced Hubble rate and the scalar field, for three
scenarios with different values of the spatial curvature.

\subsection{Constrained gravity with a matter fluid}

Having established that, qualitatively, spatial curvature does nothing
to our model, let us continue with the analysis of the field equations
in the flat case, but now in the presence of matter. We will show in
what follows that since the introduction of matter enlarges the
dimensionality of the phase space, it similarly allows the
trajectories in the contracting and expanding branches to be joined --
whatever the amount of matter present.  

If $\rho$ is the matter energy density, 
then the system will acquire one more degree
of freedom, that is, our dynamical equations are
Eqs.~(\ref{phi_0dim}), (\ref{h_0dim}) and
(\ref{rho_0dim}), which take the form
\begin{eqnarray}
 \frac{\dd h}{\dd\tau} & = & -h^{2}
 +\frac{\varphi^{2}}{2\left(1+\varphi^{6}\right)^{3/2}},
\label{h_mat}\\
 \frac{\dd\varphi}{\dd\tau} & = & -\frac{1}{2} h\left(1+4\varphi\right)+
 \frac{\varphi^{3}}{6h\sqrt{1+\varphi^{6}}}+\frac{\varepsilon}{2h},
\label{varphi_mat}\\
 \frac{\dd\varepsilon}{\dd\tau} & = & -3h\left(1+w\right)\varepsilon,
\label{matter}
\end{eqnarray}
where, as usual, $w\equiv p/\rho=\mathcal{P}/\varepsilon$ is the
equation of state for the fluid ($w=0$ for dust, $w=\frac13$ for
radiation, and $w=-\frac13$ for a curvature-like matter.)

\begin{figure}[ht]
\hskip-5mm\includegraphics[scale=0.4]{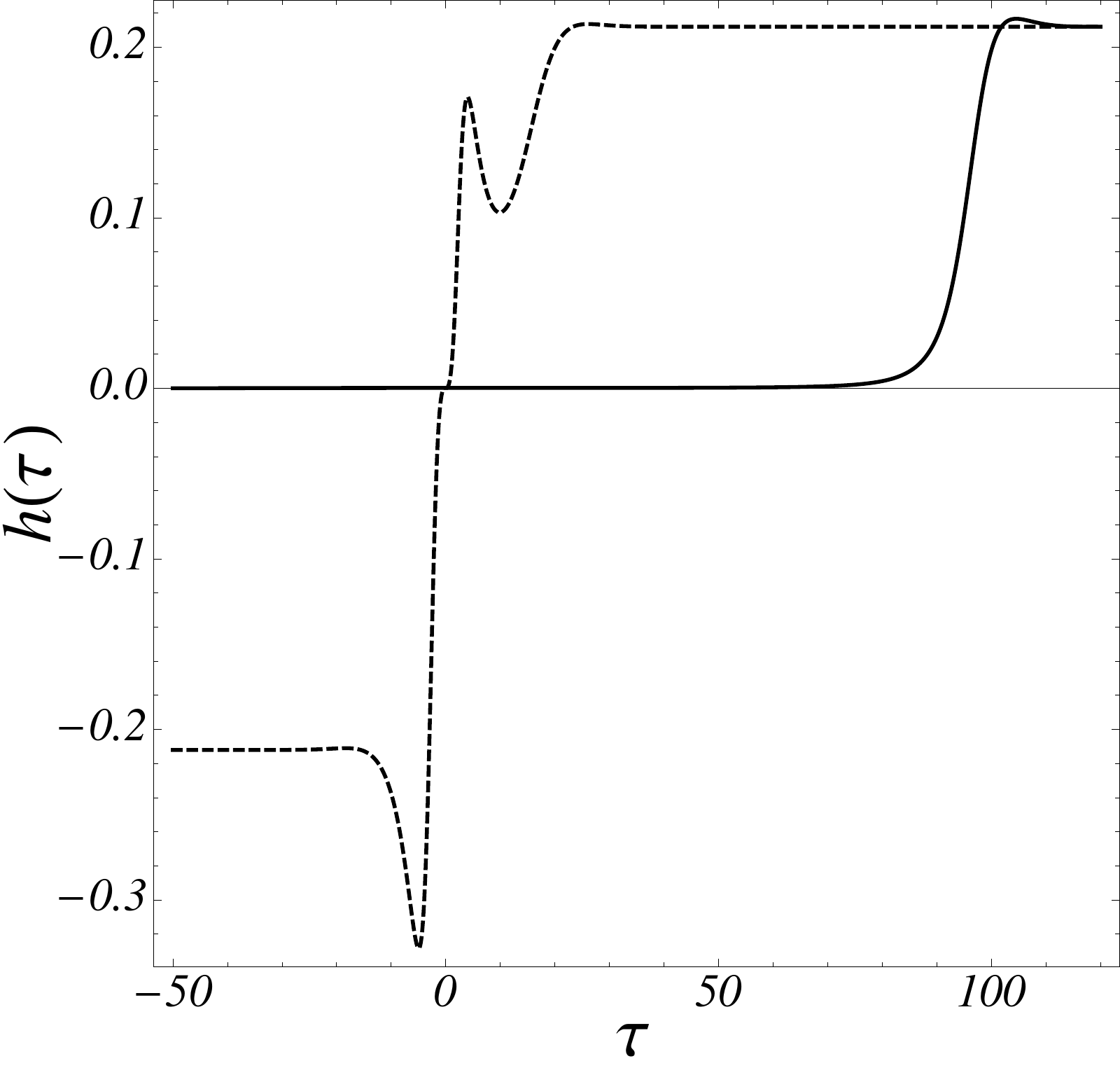}
\caption{Time evolution of the Hubble rate in vacuum (full line) and
 with matter (dashed). When a small quantity of radiation is added to
 the system [$\rho_\mathrm{rad}\left(0\right)=10^{-5}$ in this
 example], the scale factor evolves from a contracting de Sitter
 phase to the bounce. The final state corresponds to an
 ever-expanding de Sitter period. The results are exactly similar
 when, instead of radiation, one considers dust or a curvature-like
 fluid. The three cases are shown in Fig. \ref{fig:rho_t}. It is
 worth noticing that the presence of the fluid dramatically reduces
 the bounce duration so that an emergent universe model in this case
 is very unlikely.
 \label{fig:emergent_universe}}
\end{figure}

Before proceeding with the detailed analysis of the phase space of
this dynamical system, let us concentrate on a particular class of
solutions of the system (\ref{h_mat})-(\ref{matter}) which gives us
some information on how matter affects the bounce. For this purpose,
let us return to the previous vacuum case, Eqs.~(\ref{h_eq}) and
(\ref{varphi_eq}).  We have seen that, in our model, not only $h
\rightarrow 0$ at the bounce, but also that $\dd h/\dd\tau$ and
$\dd\varphi/\dd\tau$ vanish there.  The evolution at the bounce is
therefore dominated by the higher time derivatives -- which can be
tuned, through the initial conditions, to small values, making the
duration of the bounce very large.  This means that, depending on the
initial conditions of the Universe, the spacetime can stay in this
quasi-Minkowski ($h=0$) state for a long period of cosmic time before
undergoing a phase of acceleration with $\dd h/\dd\tau>0$.  However,
we see that in the presence of matter the Universe starts in a de
Sitter contracting phase and quickly evolves towards the bounce,
emerging in a super-inflationary period before reaching the
ever-expanding de Sitter phase.  This is exactly similar to what we
found for the curvature case and exemplified in
Fig. \ref{fig:emergent_universe}.

\begin{figure}[ht]
\hskip-5mm\includegraphics[scale=0.4]{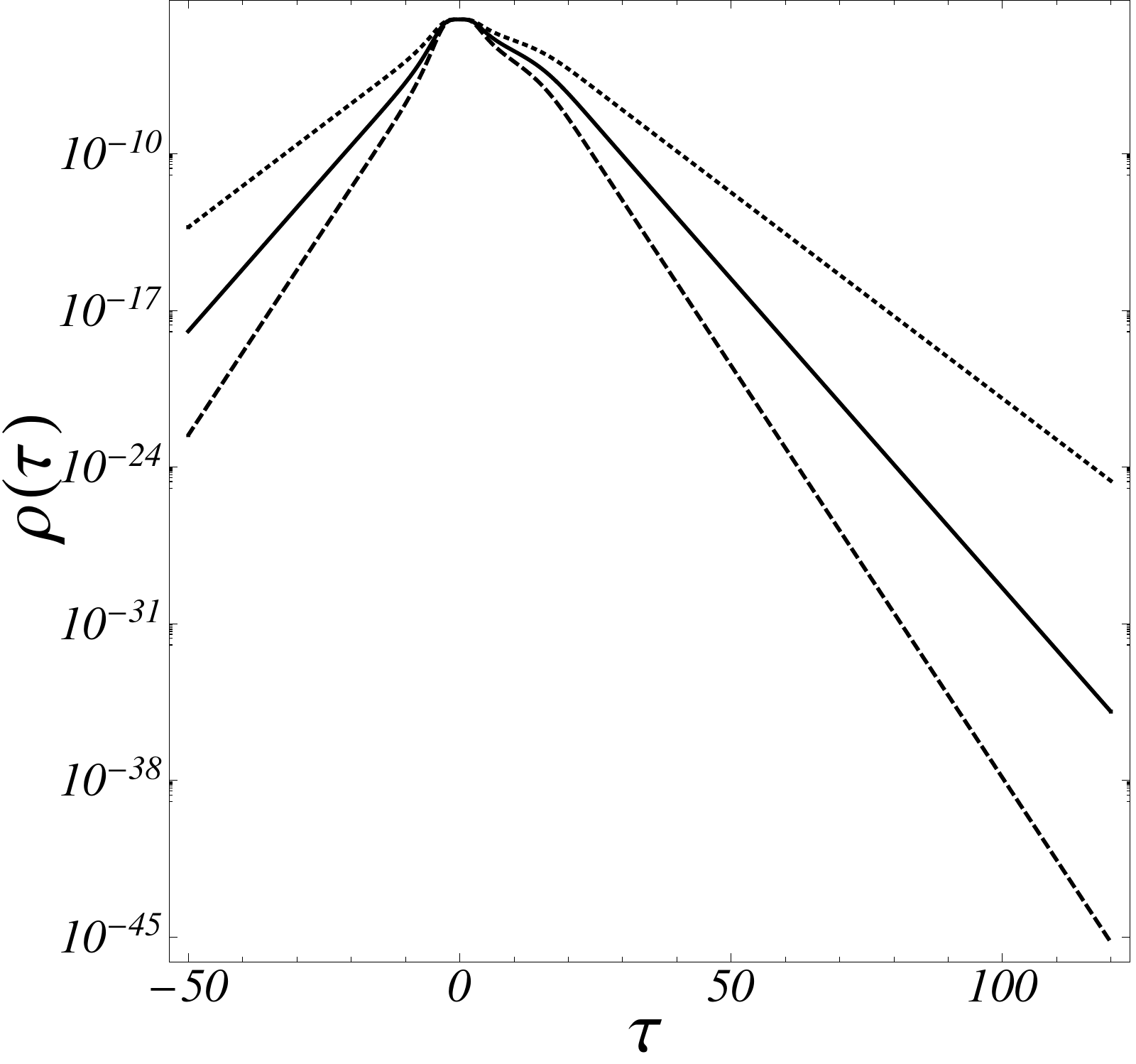}
\caption{Time evolution of the densities when the fluid consists of
 dust ($w=0$, full line), radiation ($w=\frac{1}{3}$, dashed line) and
 curvature ($w=-\frac{1}{3}$, dotted line) Hubble rate in vacuum
 (full line) and with matter (dashed). 
\label {fig:rho_t}}
\end{figure}

\begin{figure}[ht]
\hskip-5mm\includegraphics[scale=0.4]{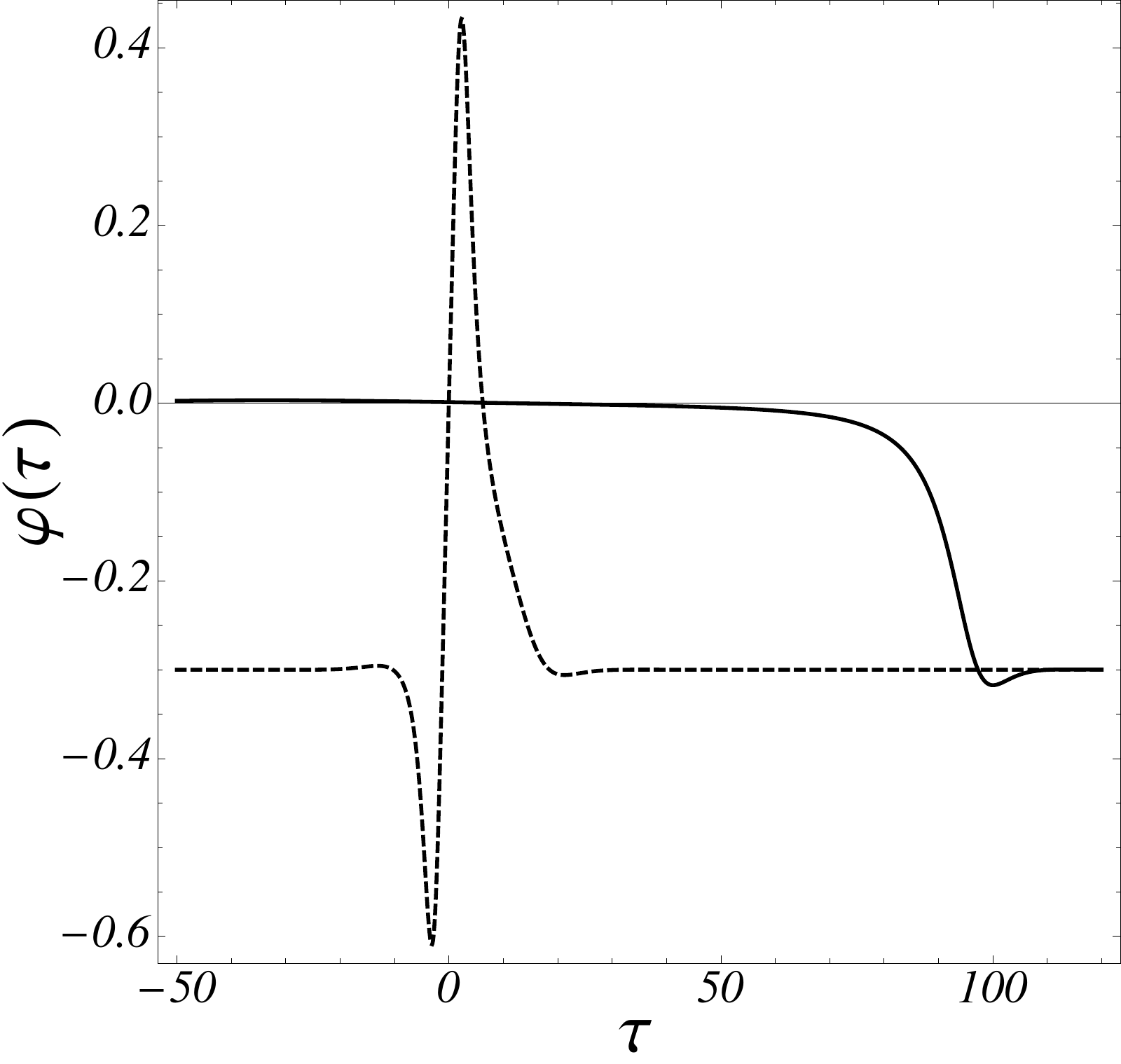}
\caption{Effect of including matter on the scalar field as a function
 of time. The full line represents the scalar field evolution in
 vacuum, while the dashed line includes a dust or radiation component
 (both curves are indistinguishable).
\label{fig:phi_tau_rad}}
\end{figure}

Notice that, as mentioned before, with matter we do not have a phase
plane anymore -- the phase space has three dimensions. The
possible trajectories that can be achieved in the projected phase
space $(\varphi,h)$ are essentially the same as those obtained in the
2 dimensional case of Fig. (\ref{fig:PS-diagram-vacuum}).

Thus we have been able to set up the cosmology for a homogeneous,
spatially flat and isotropic universe filled uniquely with dust and
radiation and that behaves as follows: the universe starts contracting
until it reaches a minimum radius, when it bounces and expands
afterwards.  The expansion period right after the bounce corresponds
to a super-inflationary period ($\dd h/\dd\tau>0$).  When the Hubble
parameter reaches a certain maximum value (which depends on the
details of the spacetime constraints), it starts to decrease,
similarly as in the usual FLRW universe, until it reaches either a de
Sitter attractor or some other regime as its final state.

\begin{figure}[ht]
\hskip-5mm\includegraphics[scale=0.5]{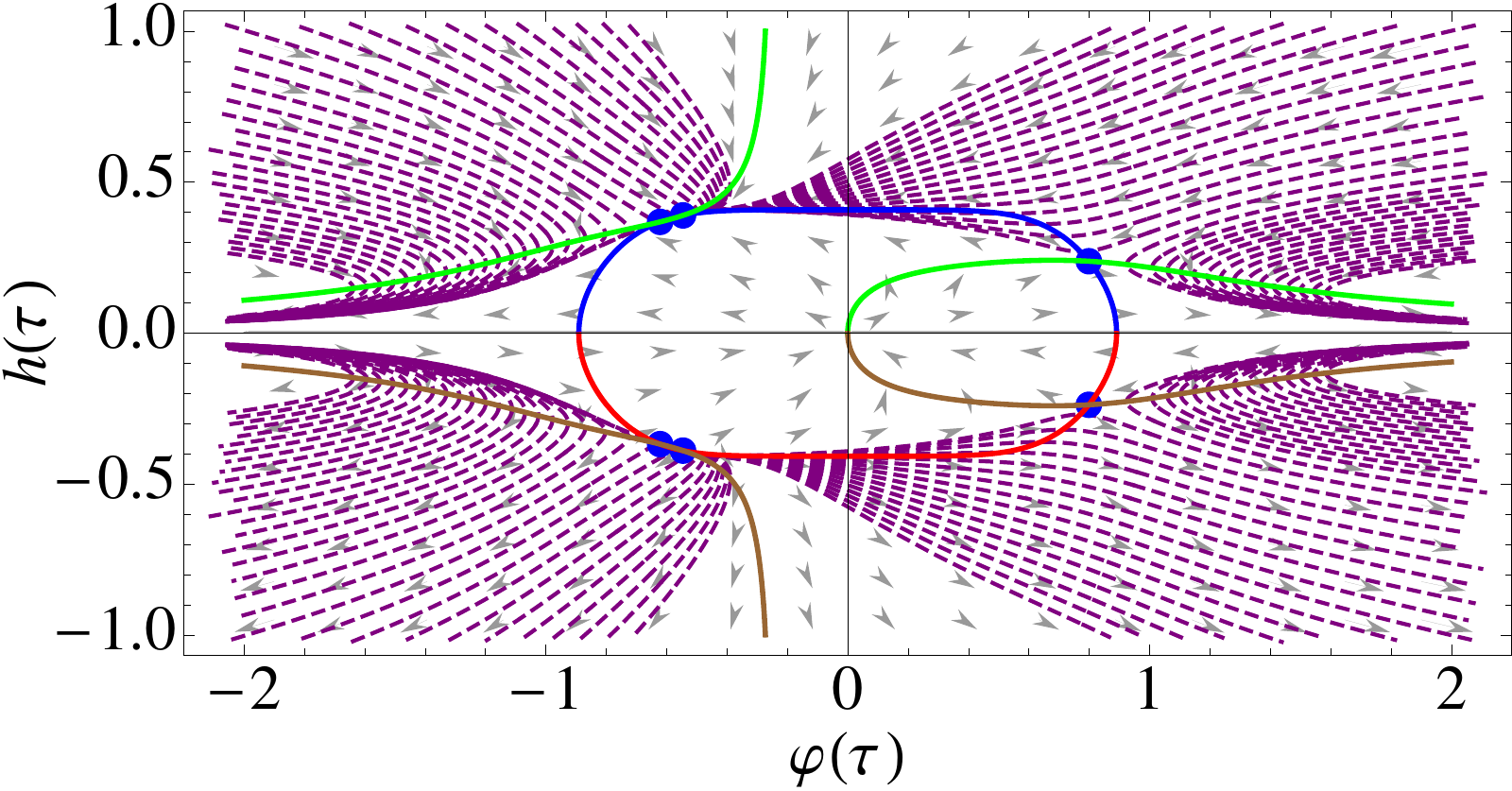}
\caption{Phase diagram for the potential $V_2$ given by
 Eq.~(\ref{invariant2}) displayed on Fig. \ref{fig:V2_phi}.
\label{fig:Phase_V2}}
\end{figure}

\begin{figure}[ht]
\hskip-5mm\includegraphics[scale=0.4]{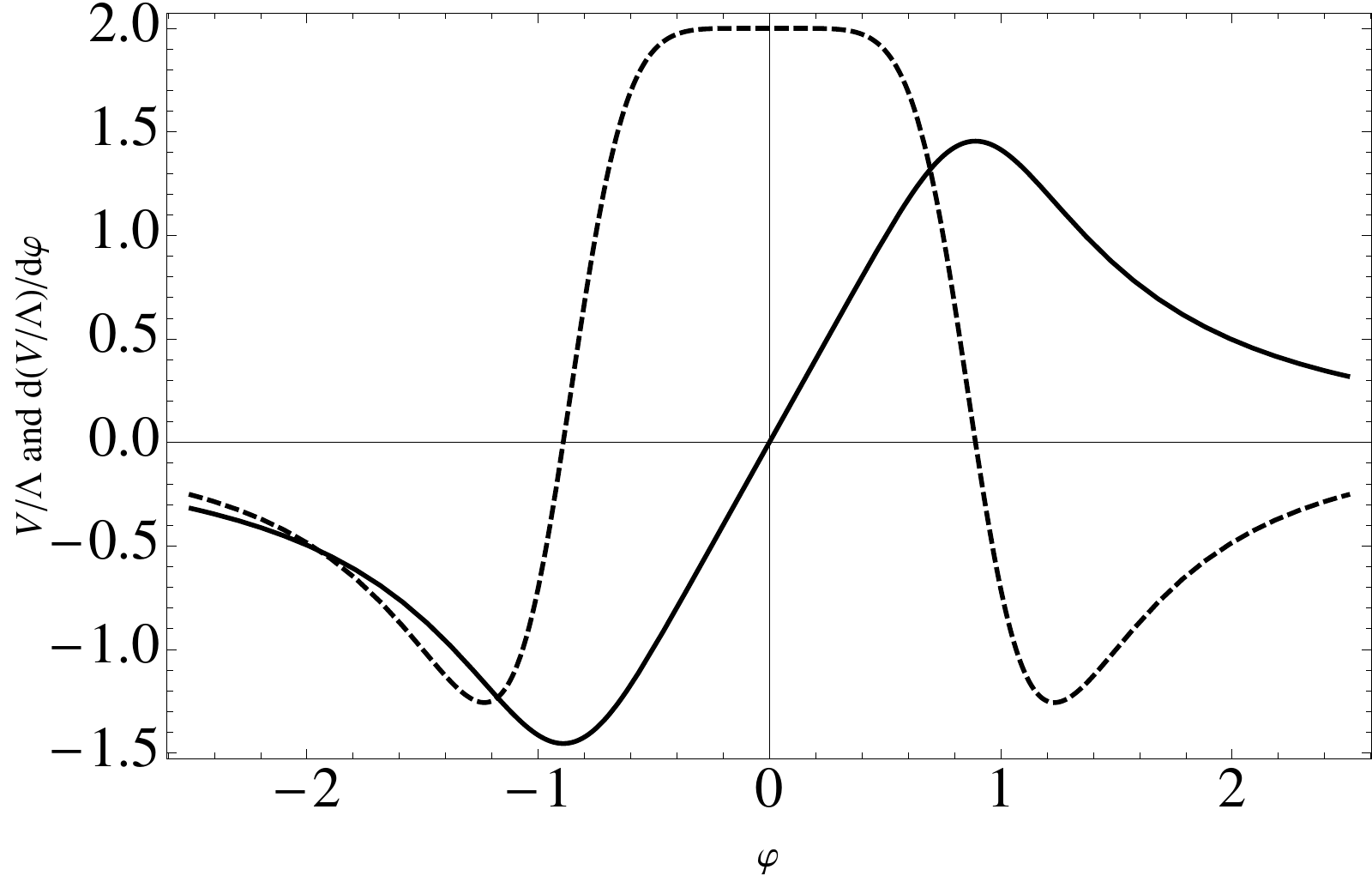}
\caption{The potential $V_2(\varphi)$ (full line) and its derivative
 (dashed) used to derive the phase portrait of
 Fig. \ref{fig:Phase_V2}.
\label{fig:V2_phi}}
\end{figure}

The behaviors discussed above are generic and do not depend crucially
on the choice (\ref{invariant}) for the potential. To see this, we
considered many different potentials, and, as a matter of example, we
show on Fig.~\ref{fig:Phase_V2} the phase diagram obtained for the
choice
\begin{equation}
 V_2\left(\varphi\right) = \frac{2\Lambda\varphi}{\sqrt{1+\varphi^{6}}},
\label{invariant2}
\end{equation}
which is represented on Fig.~\ref{fig:V2_phi}.

\subsection{Bounce Scale}

In this subsection our aim is to calculate the characteristic time
scale of the bounce described by the above model. More specifically,
we will show that, in contrast to the general relativistic description
\cite{Abramo-Peter,Martin-Peter}, the typical bounce time scale can be
made arbitrarily small in this category of theories.

Let us start with the following expansion \cite{Martin-Peter} for the
scale factor around the bounce
\begin{equation}
 a\left(\eta\right) = a_{0} \left[ 1 + \frac{1}{2} 
   \left( \frac{\eta}{\eta_{0}} \right)^{2} + \delta \left(
     \frac{\eta}{\eta_{0}} 
   \right)^{3} + \frac{5}{24} \left( 1+\xi\right) 
   \left(\frac{\eta}{\eta_{0}} \right)^{4} \right],
\label{scale_factor_expansion}
\end{equation}
where we have set the bounce time as the origin of conformal time,
i.e., $\eta_\mathrm{bounce}=0$. In Eq.~(\ref{scale_factor_expansion}),
$\eta_{0}$ provides the bounce characteristic time scale.  Similarly,
any other quantity with $0$ as a subscript means that this quantity is
to be evaluated when the bounce occurs, i.e. for $\eta=0$. As
explained in Ref.~\cite{Martin-Peter}, the parameters $\delta$ and
$\xi$ control the amplitude of the deviation from the de Sitter-like
bounce, whose scale factor is given by
$a\left(\eta\right)=a_{0}\sqrt{1+\tan^{2}\left(\eta/\eta_{0}\right)}$,
thereby explaining the factor $5/24$ in front of the fourth term in
Eq. ~(\ref{scale_factor_expansion}). 

In a similar way as for the scale factor, we assume that the scalar
field $\varphi\left(\eta\right)$, its potential $V\left(\eta\right)$
and the energy density $\rho\left(\eta\right)$ all have a Taylor
expansion about $\eta=0$, that is we set
\begin{eqnarray}
 \varphi\left(\eta\right) & = & 
 \varphi_{0}+\varphi_{1}\left(\frac{\eta}{\eta_{0}}\right)+
 \varphi_{2}\left(\frac{\eta}{\eta_{0}}\right)^{2} +
 \varphi_{3}\left(\frac{\eta}{\eta_{0}}\right)^{3}+\cdots
\label{expansion_phi}\\
 V\left(\eta\right) & = &
 V_{0}+V_{1}\left(\frac{\eta}{\eta_{0}}\right)
 +V_{2}\left(\frac{\eta}{\eta_{0}}\right)^{2} +
 V_{3}\left(\frac{\eta}{\eta_{0}}\right)^{3}+\cdots \label{expansion_V}\\
 \rho\left(\eta\right) & = & \rho_{0}+
 \rho_{1}\left(\frac{\eta}{\eta_{0}}\right)+
 \rho_{2}\left(\frac{\eta}{\eta_{0}}\right)^{2}+
 \rho_{3} \left( \frac{\eta}{\eta_{0}}\right)^{3}
 + \cdots,
 \label{expansion_rho}
\end{eqnarray}
where the parameters $\varphi_i$, $V_i$ and $\rho_i$ are to be
determined below.

In order to determine the defining parameters of the scale factor, and
hence the bounce scale itself, it is sufficient to verify in what
manner they are constrained by the field equations.  In practice, we
insert the expansion (\ref{scale_factor_expansion}) into
Eqs.~(\ref{VincConf}), (\ref{FriedConf}) and (\ref{ConsConf}) and
identify terms order by order in $\eta/\eta_0\ll 1$. This leads to
\begin{eqnarray}
 \rho_1&=&0,\label{rho1}\\
 \rho_2&=&-\frac32 (1+w)\rho_0,\label{rho2}\\
 \rho_3&=&-3\delta (1+w)\rho_0,\label{rho3}
\end{eqnarray}
which expresses the first terms in the energy density. Note at this
point that a symmetric bounce, having $\delta=0$, leads to an even
behavior for the density, as expected. The first equality merely
expresses energy conservation at the bounce conformal time.

The cosmic time bounce duration $t_0$ can then be evaluated as
\begin{equation}
 t_0^2\equiv a_{0}^{2}\eta_{0}^{2} = 
 \frac{12\varphi_{1}}{V_{1}}=12\left(\left. 
     \frac{\dd V}{\dd\varphi}\right|_{0}\right)^{-1},
\label{bounce_scale}
\end{equation}
where the last term is the derivative of the potential with respect to
$\varphi$ evaluated at the bounce conformal time $\eta=0$; this
remains a free parameter, as can be seen from inspection of the
following relations
\begin{eqnarray}
 V_{0} & = & \frac{6\Ka}{a_0^2}-16\pi\GN \rho_{0},\label{V0}\\
 V_{2}& = & \left. \frac{\dd V}{\dd\varphi}\right|_{0} \left(3\delta
   \varphi_1 +\varphi_2\right),\label{V2_0}
\end{eqnarray}
relating the parameters for the scalar field potential. Note at this
point that for negative or vanishing curvature $\Ka$, the potential
needs to be negative at the bounce, so that, as in GR, a simple
massive scalar field cannot lead to a bouncing phase unless the
Universe is closed.

At the next orders, we obtain relationship between the
higher derivatives of the scalar field at the bounce and the energy
density at that point as well as the asymmetry of the bounce, namely
\begin{eqnarray}
 \varphi_2 &=& -\frac12 \left(1+4\varphi_0\right)+6 
 \left(\left.  \frac{\dd V}{\dd\varphi}\right|_{0}\right)^{-1}
 \left[\frac{\Ka}{a_0^2}-4\pi\GN \rho_0\left(1+w\right)
 \right],\cr&&\label{phi2}\\
 \varphi_3&=&-\delta\left(1+4\varphi_0\right) -\frac23 \varphi_1,\label{phi3}
\end{eqnarray}
and finally one can identify the fourth order term in the scale
factor expansion $\xi$ as
\begin{eqnarray}
 \xi &=& \frac65\left\{ \frac{1}{V_1} \left[
     V_3+24\pi\GN \delta\rho_0\left(1+w\right)
     -24\delta\frac{\Ka}{a_0^2}\right]\right.\cr
   &&\hskip2cm+\left.\frac{2\delta\left(1+4\varphi_0\right)-\varphi_3}{\varphi_1}
 \right\}.
\end{eqnarray}

An important conclusion that can be drawn from these simple relations
is that the bounce scale can be made as small as one wants. Clearly,
we see that the larger the derivative of the potential at the bounce
or, put in another way, the larger the constant $\sqrt{\Lambda}$ which
defines the dimension of $V\left(\varphi\right)$ is, the smaller the
bounce characteristic conformal time scale will be, for a given bounce
length scale $a_0$. This result can be consistently checked by
numerical integration of system (\ref{FriedConf}): consider the vacuum
equations of the last subsection, Eqs.~(\ref{h_eq}) and
(\ref{varphi_eq}). At the point $\left(\varphi=0,h=0\right)$, we
recall that the derivative of the potential is exactly zero. But,
according to the relation (\ref{bounce_scale}), this implies that the
physical time duration $t_{0}=a_{0}\eta_{0}$ of the bounce goes to
infinity or, put in another way, that there is no bounce at all. This
confirms the results for vacuum obtained in the last section.

\section{Conclusion}

Cosmological inflation now probably deserves to be included in
the standard picture of the Universe. 
However, as inflation cannot be directly observed, 
it is desirable to find
other, challenging models that could similarly solve the problems of
non inflationary cosmology while at the same time making different
predictions for data sets -- some of which may yet be observed 
as, e.g., non gaussianity or tensor modes of perturbations. Such a 
non-inflationary model could include a phase of contraction 
followed by a bounce, leading to the current expansion.

Having a bouncing phase in the early Universe is a highly non-trivial
demand in the framework of general relativity with well-behaved matter
content. It usually requires a strictly positive value for the spatial
curvature, in contradiction with the current observational
data. Therefore, unless a phase of inflation is invoked after the
bounce occurred, either the matter content or the gravity theory must
be changed. 

The simplest way for a bounce to take place in a regular matter theory
consists in demanding a positive spatial curvature to compensate
for the positive energy density of matter, allowing the Hubble parameter
to vanish at the bounce. 
But by doing so, it was found that the time duration of the bounce
was bounded from below so that an arbitrary short bounce could not
take place. It was suggested that this was entirely due to the
curvature, whose value was indeed crucial in limiting the bounce
duration.

In this paper our main purpose was to investigate a non singular
bounce in the framework of the modified gravity model of
\cite{Mukhanov-Brandenberger-PRD}.  We found solutions for which the
universe is described by a period of contraction followed by an
expanding, super-inflationary phase before connecting to a more usual
radiation- or matter-dominated epoch of the FLRW universe. The final
period then emerges as an ever-expanding de Sitter universe. We have
computed the duration of the bounce for that model, and we found that,
whatever the bounce, irrespectively of whether or not it is symmetric, 
whether or not there is spatial curvature 
(independently of the sign of this curvature), and
whether in vacuum or not, its typical duration is unconstrained.

This arbitrariness in the duration of the bounce
could have important consequences for the evolution of fluctuations
through the bounce. However, although the background models are
ruled by second-order equations, perturbations will almost surely
show signs of the underlying instability of our higher-order
gravity model, so at this stage it is not clear that
perturbations in the framework of our models could be sensible.

\subsection*{Acknowledgements}

R.~A. and I.~Y. would like to thank the Institut d'Astrophysique de
Paris, and P.~P., the Instituto de F\'{\i}sica da Universidade de 
S\~ao Paulo, for their warm hospitality. We also would like to thank 
FAPESP, CNPq and CAPES of Brazil, and COFECUB of France, for financial 
support.

\bibliography{biblio}

\begin{thebibliography}{33}
\expandafter\ifx\csname natexlab\endcsname\relax\def\natexlab#1{#1}\fi
\expandafter\ifx\csname bibnamefont\endcsname\relax
  \def\bibnamefont#1{#1}\fi
\expandafter\ifx\csname bibfnamefont\endcsname\relax
  \def\bibfnamefont#1{#1}\fi
\expandafter\ifx\csname citenamefont\endcsname\relax
  \def\citenamefont#1{#1}\fi
\expandafter\ifx\csname url\endcsname\relax
  \def\url#1{\texttt{#1}}\fi
\expandafter\ifx\csname urlprefix\endcsname\relax\def\urlprefix{URL }\fi
\providecommand{\bibinfo}[2]{#2}
\providecommand{\eprint}[2][]{\url{#2}}

\bibitem[{\citenamefont{{Komatsu {\sl et al}}}(2009)}]{Komatsu:2008hk}
\bibinfo{author}{\bibfnamefont{E.}~\bibnamefont{{Komatsu {\sl et al}}}},
  \bibinfo{journal}{Ap. J. Supp.} \textbf{\bibinfo{volume}{180}},
  \bibinfo{pages}{330} (\bibinfo{year}{2009}), \eprint{0803.0547}.

\bibitem[{\citenamefont{Guth}(1981)}]{Guth}
\bibinfo{author}{\bibfnamefont{A.~H.} \bibnamefont{Guth}},
  \bibinfo{journal}{Phys. Rev. D} \textbf{\bibinfo{volume}{23}},
  \bibinfo{pages}{347} (\bibinfo{year}{1981}).

\bibitem[{\citenamefont{Linde}(1982)}]{Linde-PL}
\bibinfo{author}{\bibfnamefont{A.~D.} \bibnamefont{Linde}},
  \bibinfo{journal}{Phys. Lett.} \textbf{\bibinfo{volume}{B108}},
  \bibinfo{pages}{389} (\bibinfo{year}{1982}).

\bibitem[{\citenamefont{Albrecht and Steinhardt}(1982)}]{Albrecht-Steinhardt}
\bibinfo{author}{\bibfnamefont{A.}~\bibnamefont{Albrecht}} \bibnamefont{and}
  \bibinfo{author}{\bibfnamefont{P.~J.} \bibnamefont{Steinhardt}},
  \bibinfo{journal}{Phys. Rev. Lett.} \textbf{\bibinfo{volume}{48}},
  \bibinfo{pages}{1220} (\bibinfo{year}{1982}).

\bibitem[{\citenamefont{Borde and Vilenkin}(1997)}]{Borde-Vilenkin-PRD}
\bibinfo{author}{\bibfnamefont{A.}~\bibnamefont{Borde}} \bibnamefont{and}
  \bibinfo{author}{\bibfnamefont{A.}~\bibnamefont{Vilenkin}},
  \bibinfo{journal}{Phys. Rev. D} \textbf{\bibinfo{volume}{56}},
  \bibinfo{pages}{717} (\bibinfo{year}{1997}).

\bibitem[{\citenamefont{Borde and Vilenkin}(1994)}]{Borde-Vilenkin-PRL}
\bibinfo{author}{\bibfnamefont{A.}~\bibnamefont{Borde}} \bibnamefont{and}
  \bibinfo{author}{\bibfnamefont{A.}~\bibnamefont{Vilenkin}},
  \bibinfo{journal}{Phys. Rev. Lett.} \textbf{\bibinfo{volume}{72}},
  \bibinfo{pages}{3305} (\bibinfo{year}{1994}).

\bibitem[{\citenamefont{Borde et~al.}(2003)\citenamefont{Borde, Guth, and
  Vilenkin}}]{Borde-Guth-Vilenkin-PRL}
\bibinfo{author}{\bibfnamefont{A.}~\bibnamefont{Borde}},
  \bibinfo{author}{\bibfnamefont{A.~H.} \bibnamefont{Guth}}, \bibnamefont{and}
  \bibinfo{author}{\bibfnamefont{A.}~\bibnamefont{Vilenkin}},
  \bibinfo{journal}{Phys. Rev. Lett.} \textbf{\bibinfo{volume}{90}},
  \bibinfo{pages}{151301} (\bibinfo{year}{2003}).

\bibitem[{\citenamefont{Gasperini et~al.}(2003)\citenamefont{Gasperini,
  Giovannini, and Veneziano}}]{Gasperini-Veneziano-PR}
\bibinfo{author}{\bibfnamefont{M.}~\bibnamefont{Gasperini}},
  \bibinfo{author}{\bibfnamefont{M.}~\bibnamefont{Giovannini}},
  \bibnamefont{and}
  \bibinfo{author}{\bibfnamefont{G.}~\bibnamefont{Veneziano}},
  \bibinfo{journal}{Phys. Rep.} \textbf{\bibinfo{volume}{373}},
  \bibinfo{pages}{1} (\bibinfo{year}{2003}).

\bibitem[{\citenamefont{Lidsey et~al.}(2000)\citenamefont{Lidsey, Wands, and
  Copeland}}]{Lidsey-PR}
\bibinfo{author}{\bibfnamefont{J.~E.} \bibnamefont{Lidsey}},
  \bibinfo{author}{\bibfnamefont{D.}~\bibnamefont{Wands}}, \bibnamefont{and}
  \bibinfo{author}{\bibfnamefont{E.~J.} \bibnamefont{Copeland}},
  \bibinfo{journal}{Physics Reports} \textbf{\bibinfo{volume}{337}},
  \bibinfo{pages}{343 } (\bibinfo{year}{2000}).

\bibitem[{\citenamefont{Murphy}(1973)}]{Murphy}
\bibinfo{author}{\bibfnamefont{G.~L.} \bibnamefont{Murphy}},
  \bibinfo{journal}{Phys. Rev. D} \textbf{\bibinfo{volume}{8}},
  \bibinfo{pages}{4231} (\bibinfo{year}{1973}).

\bibitem[{\citenamefont{Melnikov and Orlov}(1979)}]{Melnikov}
\bibinfo{author}{\bibfnamefont{V.~N.} \bibnamefont{Melnikov}} \bibnamefont{and}
  \bibinfo{author}{\bibfnamefont{S.~V.} \bibnamefont{Orlov}},
  \bibinfo{journal}{Phys. Lett. A} \textbf{\bibinfo{volume}{70}},
  \bibinfo{pages}{263 } (\bibinfo{year}{1979}).

\bibitem[{\citenamefont{Fabris et~al.}(2003)\citenamefont{Fabris, Furtado,
  Peter, and Pinto-Neto}}]{Fabris-Peter}
\bibinfo{author}{\bibfnamefont{J.~C.} \bibnamefont{Fabris}},
  \bibinfo{author}{\bibfnamefont{R.~G.} \bibnamefont{Furtado}},
  \bibinfo{author}{\bibfnamefont{P.}~\bibnamefont{Peter}}, \bibnamefont{and}
  \bibinfo{author}{\bibfnamefont{N.}~\bibnamefont{Pinto-Neto}},
  \bibinfo{journal}{Phys. Rev. D} \textbf{\bibinfo{volume}{67}},
  \bibinfo{pages}{124003} (\bibinfo{year}{2003}).

\bibitem[{\citenamefont{Martin and Peter}(2003)}]{Martin-Peter}
\bibinfo{author}{\bibfnamefont{J.}~\bibnamefont{Martin}} \bibnamefont{and}
  \bibinfo{author}{\bibfnamefont{P.}~\bibnamefont{Peter}},
  \bibinfo{journal}{Phys. Rev.} \textbf{\bibinfo{volume}{D68}},
  \bibinfo{pages}{103517} (\bibinfo{year}{2003}), \eprint{hep-th/0307077}.

\bibitem[{\citenamefont{Martin et~al.}(2002)\citenamefont{Martin, Peter,
  Pinto~Neto, and Schwarz}}]{Martin-Peter-Pinto-Neto-PRD}
\bibinfo{author}{\bibfnamefont{J.}~\bibnamefont{Martin}},
  \bibinfo{author}{\bibfnamefont{P.}~\bibnamefont{Peter}},
  \bibinfo{author}{\bibfnamefont{N.}~\bibnamefont{Pinto~Neto}},
  \bibnamefont{and} \bibinfo{author}{\bibfnamefont{D.~J.}
  \bibnamefont{Schwarz}}, \bibinfo{journal}{Phys. Rev.}
  \textbf{\bibinfo{volume}{D65}}, \bibinfo{pages}{123513}
  (\bibinfo{year}{2002}), \eprint{hep-th/0112128}.

\bibitem[{\citenamefont{Veneziano}(1997)}]{Veneziano-arxiv}
\bibinfo{author}{\bibfnamefont{G.}~\bibnamefont{Veneziano}}
  (\bibinfo{year}{1997}), \eprint{hep-th/9802057}.

\bibitem[{\citenamefont{Brustein and Veneziano}(1994)}]{Brustein-Veneziano}
\bibinfo{author}{\bibfnamefont{R.}~\bibnamefont{Brustein}} \bibnamefont{and}
  \bibinfo{author}{\bibfnamefont{G.}~\bibnamefont{Veneziano}},
  \bibinfo{journal}{Phys. Lett.} \textbf{\bibinfo{volume}{B329}},
  \bibinfo{pages}{429} (\bibinfo{year}{1994}), \eprint{hep-th/9403060}.

\bibitem[{\citenamefont{Finelli and
  Brandenberger}(1999)}]{finelli-brandenberger-PRL}
\bibinfo{author}{\bibfnamefont{F.}~\bibnamefont{Finelli}} \bibnamefont{and}
  \bibinfo{author}{\bibfnamefont{R.}~\bibnamefont{Brandenberger}},
  \bibinfo{journal}{Phys. Rev. Lett.} \textbf{\bibinfo{volume}{82}},
  \bibinfo{pages}{1362} (\bibinfo{year}{1999}).

\bibitem[{\citenamefont{Mukhanov et~al.}(1992)\citenamefont{Mukhanov, Feldman,
  and Brandenberger}}]{Mukhanov-Brandenberger-Perturbations}
\bibinfo{author}{\bibfnamefont{V.~F.} \bibnamefont{Mukhanov}},
  \bibinfo{author}{\bibfnamefont{H.~A.} \bibnamefont{Feldman}},
  \bibnamefont{and} \bibinfo{author}{\bibfnamefont{R.~H.}
  \bibnamefont{Brandenberger}}, \bibinfo{journal}{Phys. Rept.}
  \textbf{\bibinfo{volume}{215}}, \bibinfo{pages}{203} (\bibinfo{year}{1992}).

\bibitem[{\citenamefont{Durrer and Vernizzi}(2002)}]{Durrer}
\bibinfo{author}{\bibfnamefont{R.}~\bibnamefont{Durrer}} \bibnamefont{and}
  \bibinfo{author}{\bibfnamefont{F.}~\bibnamefont{Vernizzi}},
  \bibinfo{journal}{Phys. Rev. D} \textbf{\bibinfo{volume}{66}},
  \bibinfo{pages}{083503} (\bibinfo{year}{2002}).

\bibitem[{\citenamefont{Martin and Peter}(2004)}]{Martin-Peter-PRL}
\bibinfo{author}{\bibfnamefont{J.}~\bibnamefont{Martin}} \bibnamefont{and}
  \bibinfo{author}{\bibfnamefont{P.}~\bibnamefont{Peter}},
  \bibinfo{journal}{Phys. Rev. Lett.} \textbf{\bibinfo{volume}{92}},
  \bibinfo{pages}{061301} (\bibinfo{year}{2004}).

\bibitem[{\citenamefont{Armendariz-Picon
  et~al.}(2001)\citenamefont{Armendariz-Picon, Mukhanov, and
  Steinhardt}}]{Armendariz-PRD}
\bibinfo{author}{\bibfnamefont{C.}~\bibnamefont{Armendariz-Picon}},
  \bibinfo{author}{\bibfnamefont{V.~F.} \bibnamefont{Mukhanov}},
  \bibnamefont{and} \bibinfo{author}{\bibfnamefont{P.~J.}
  \bibnamefont{Steinhardt}}, \bibinfo{journal}{Phys. Rev.}
  \textbf{\bibinfo{volume}{D63}}, \bibinfo{pages}{103510}
  (\bibinfo{year}{2001}), \eprint{astro-ph/0006373}.

\bibitem[{\citenamefont{Abramo and Peter}(2007)}]{Abramo-Peter}
\bibinfo{author}{\bibfnamefont{L.~R.} \bibnamefont{Abramo}} \bibnamefont{and}
  \bibinfo{author}{\bibfnamefont{P.}~\bibnamefont{Peter}},
  \bibinfo{journal}{JCAP} \textbf{\bibinfo{volume}{0709}}, \bibinfo{pages}{001}
  (\bibinfo{year}{2007}), \eprint{arXiv:0705.2893 [astro-ph]}.

\bibitem[{\citenamefont{Mukhanov and
  Brandenberger}(1992)}]{Mukhanov-Brandenberger-PRL}
\bibinfo{author}{\bibfnamefont{V.~F.} \bibnamefont{Mukhanov}} \bibnamefont{and}
  \bibinfo{author}{\bibfnamefont{R.~H.} \bibnamefont{Brandenberger}},
  \bibinfo{journal}{Phys. Rev. Lett.} \textbf{\bibinfo{volume}{68}},
  \bibinfo{pages}{1969} (\bibinfo{year}{1992}).

\bibitem[{\citenamefont{Brandenberger et~al.}(1993)\citenamefont{Brandenberger,
  Mukhanov, and Sornborger}}]{Mukhanov-Brandenberger-PRD}
\bibinfo{author}{\bibfnamefont{R.}~\bibnamefont{Brandenberger}},
  \bibinfo{author}{\bibfnamefont{V.}~\bibnamefont{Mukhanov}}, \bibnamefont{and}
  \bibinfo{author}{\bibfnamefont{A.}~\bibnamefont{Sornborger}},
  \bibinfo{journal}{Phys. Rev. D} \textbf{\bibinfo{volume}{48}},
  \bibinfo{pages}{1629} (\bibinfo{year}{1993}).

\bibitem[{\citenamefont{Brandenberger et~al.}(1998)\citenamefont{Brandenberger,
  Easther, and Maia}}]{Brandenberger-Easther}
\bibinfo{author}{\bibfnamefont{R.~H.} \bibnamefont{Brandenberger}},
  \bibinfo{author}{\bibfnamefont{R.}~\bibnamefont{Easther}}, \bibnamefont{and}
  \bibinfo{author}{\bibfnamefont{J.}~\bibnamefont{Maia}},
  \bibinfo{journal}{JHEP} \textbf{\bibinfo{volume}{08}}, \bibinfo{pages}{007}
  (\bibinfo{year}{1998}), \eprint{gr-qc/9806111}.

\bibitem[{\citenamefont{'t~Hooft and Veltman}(1974)}]{Hooft}
\bibinfo{author}{\bibfnamefont{G.}~\bibnamefont{'t~Hooft}} \bibnamefont{and}
  \bibinfo{author}{\bibfnamefont{M.~J.~G.} \bibnamefont{Veltman}},
  \bibinfo{journal}{Annales Inst. H. Poincaré} \textbf{\bibinfo{volume}{A20}},
  \bibinfo{pages}{69} (\bibinfo{year}{1974}).

\bibitem[{\citenamefont{Deser et~al.}(1974)\citenamefont{Deser, Tsao, and van
  Nieuwenhuizen}}]{Deser-Nieuwenhuizen-PRD}
\bibinfo{author}{\bibfnamefont{S.}~\bibnamefont{Deser}},
  \bibinfo{author}{\bibfnamefont{H.-S.} \bibnamefont{Tsao}}, \bibnamefont{and}
  \bibinfo{author}{\bibfnamefont{P.}~\bibnamefont{van Nieuwenhuizen}},
  \bibinfo{journal}{Phys. Rev. D} \textbf{\bibinfo{volume}{10}},
  \bibinfo{pages}{3337} (\bibinfo{year}{1974}).

\bibitem[{\citenamefont{Stelle}(1977)}]{Stelle}
\bibinfo{author}{\bibfnamefont{K.~S.} \bibnamefont{Stelle}},
  \bibinfo{journal}{Phys. Rev. D} \textbf{\bibinfo{volume}{16}},
  \bibinfo{pages}{953} (\bibinfo{year}{1977}).

\bibitem[{\citenamefont{Chiba}(2005)}]{Chiba-JCAP}
\bibinfo{author}{\bibfnamefont{T.}~\bibnamefont{Chiba}},
  \bibinfo{journal}{JCAP} \textbf{\bibinfo{volume}{0503}}, \bibinfo{pages}{008}
  (\bibinfo{year}{2005}), \eprint{gr-qc/0502070}.

\bibitem[{\citenamefont{Nunez and Solganik}(2005)}]{Nunez}
\bibinfo{author}{\bibfnamefont{A.}~\bibnamefont{Nunez}} \bibnamefont{and}
  \bibinfo{author}{\bibfnamefont{S.}~\bibnamefont{Solganik}},
  \bibinfo{journal}{Phys. Lett.} \textbf{\bibinfo{volume}{B608}},
  \bibinfo{pages}{189} (\bibinfo{year}{2005}), \eprint{hep-th/0411102}.

\bibitem[{\citenamefont{Woodard}(2007)}]{Woodard}
\bibinfo{author}{\bibfnamefont{R.~P.} \bibnamefont{Woodard}},
  \bibinfo{journal}{Lect. Notes Phys.} \textbf{\bibinfo{volume}{720}},
  \bibinfo{pages}{403} (\bibinfo{year}{2007}), \eprint{astro-ph/0601672}.

\bibitem[{\citenamefont{Bruneton and
  Esposito-Farese}(2007)}]{Bruneton-Esposito}
\bibinfo{author}{\bibfnamefont{J.-P.} \bibnamefont{Bruneton}} \bibnamefont{and}
  \bibinfo{author}{\bibfnamefont{G.}~\bibnamefont{Esposito-Farese}},
  \bibinfo{journal}{Phys. Rev.} \textbf{\bibinfo{volume}{D76}},
  \bibinfo{pages}{124012} (\bibinfo{year}{2007}), \eprint{0705.4043}.

\bibitem[{\citenamefont{Jeffreys and Jeffreys}(1988)}]{Jeffreys}
\bibinfo{author}{\bibfnamefont{H.}~\bibnamefont{Jeffreys}} \bibnamefont{and}
  \bibinfo{author}{\bibfnamefont{B.~S.} \bibnamefont{Jeffreys}},
  \emph{\bibinfo{title}{{Methods of Mathematical Physics, 3rd Ed.}}}
  (\bibinfo{publisher}{Cambridge, UK: Cambridge University Press},
  \bibinfo{year}{1988}).

\end{thebibliography}

\end{document}